\documentclass{jpsj3}
\usepackage{txfonts}
\usepackage{overcite,mathptm,times,graphicx}
\usepackage{subfigure}
\newcounter{theorem}
\renewcommand\thetheorem{\arabic{section}.\arabic{theorem}}

\newenvironment{lemma}{\par\medskip\noindent\begingroup{\bf Lemma
             \stepcounter{theorem}\thetheorem.}\ \itshape
             \def\@currentlabel{\thetheorem}}{\endgroup\par\medskip}
\newenvironment{theorem}{\par\medskip\noindent\begingroup{\bf Theorem
             \stepcounter{theorem}\thetheorem.}\ \itshape
             \def\@currentlabel{\thetheorem}}{\endgroup\par\medskip}

\newenvironment{remark}{\par\medskip\noindent\begingroup{\bf Remark
             \stepcounter{theorem}\thetheorem.}\
             \def\@currentlabel{\thetheorem}}{\endgroup\par\medskip}

\newenvironment{proof}{\par\noindent{\bf Proof.}
}{\proofbox\par\medskip}

\newdimen\LENB \newdimen\LENW \newdimen\THI
\newdimen\LENWH \newdimen\LENTOT \newcount\N
\def\vbrknlnele#1#2#3{
  \LENB=#1pt \LENW=#2pt \THI=#3pt
  \LENWH=\LENW \divide\LENWH by 2
  \LENTOT=\LENB \advance\LENTOT by \LENW
  \vbox to \LENTOT{
    \vbox to \LENWH{}
    \nointerlineskip
    \vbox to \LENB{\hbox to \THI{\vrule width \THI height \LENB}}
    \nointerlineskip
    \vbox to \LENWH{}
  }}

\def\vbrknln#1{
  \N=#1
  \vcenter{
    \vbox{
      \loop\ifnum\N>0
        \vbox to 4pt{\vbrknlnele{2}{2}{0.1}}
        \nointerlineskip
        \advance\N by -1
      \repeat
  }}}

\def\vbl#1{\hskip-5pt \vbrknln{#1} \hskip-5pt}

\def\hbrknlnele#1#2#3{
  \LENB=#1pt \LENW=#2pt \THI=#3pt
  \LENTOT=\LENB \advance\LENTOT by \LENW
  \vcenter{
    \vbox to \THI{
      \hbox to \LENTOT{
        \hfil
        \vrule width \LENB height \THI
        \hfil}
  }}}

\def\hblele{\hbrknlnele{2}{2.2}{0.1}}

\def\hblfil{\cleaders\hbox{$\mkern1mu \hblele \mkern1mu$}\hfill}

\title{Multi-dark soliton solutions of the two-dimensional multi-component Yajima-Oikawa
systems}

\author{Junchao Chen$^{1,2}$, Yong Chen$^1$\thanks{ychen@sei.ecnu.edu.cn}, Bao-Feng Feng$^2$\thanks{feng@utpa.edu}, Ken-ichi Maruno$^{3}$\thanks{kmaruno@waseda.jp}}
\inst{$^1$Shanghai Key Laboratory of Trustworthy Computing, East China Normal University, Shanghai, 200062, People's Republic of China \\
$^2$ Department of Mathematics, The University of Texas-Pan American, Edinburg, TX 78541, USA \\
$^3$~Department of Applied Mathematics, School of Fundamental Science
and Engineering, Faculty of Science and Engineering,
Waseda University, 3-4-1 Okubo, Shinjuku-ku, Tokyo 169-8555, Japan} 

\abst{
We present a general form of multi-dark soliton solutions of
two-dimensional multi-component soliton systems.
Multi-dark soliton solutions of the two-dimensional (2D) and one-dimensional (1D) multi-component Yajima-Oikawa
 (YO) systems, which are often called
the 2D and 1D multi-component long wave-short wave resonance interaction systems, are studied in detail.
Taking the 2D coupled YO system with two short wave
and one long wave components
as an example,
we derive the general $N$-dark-dark soliton solution in both the Gram
 type and Wronski type determinant forms for the 2D coupled
 YO system via the KP hierarchy reduction method.
By imposing certain constraint conditions, the general $N$-dark-dark soliton
 solution of the 1D coupled YO system is further obtained.
The dynamics of one dark-dark and two dark-dark solitons are analyzed in detail.
In contrast with bright-bright soliton collisions,
it is shown that dark-dark soliton collisions are elastic and there is
 no energy exchange among solitons in different components.
Moreover, the dark-dark soliton bound states including the stationary
 and moving ones are discussed.
For the stationary case, the bound states exist up to arbitrary order, whereas,
for the moving case, only the two-soliton bound state
is possible under the condition
that the coefficients of nonlinear terms have opposite signs.}

\begin{document}
\maketitle

\vspace{5cm}

\section{Introduction}

The two-dimensional (2D) coupled Yajima-Oikawa (YO) system,
or the so-called 2D coupled long wave-short wave resonance interaction system\cite{ohta2007two}:
\begin{eqnarray}
\label{nyo-01} &&\textmd{i}(S^{(1)}_t + S^{(1)}_y) - S^{(1)}_{xx}+ L S^{(1)}=0,\\
\label{nyo-02} &&\textmd{i}(S^{(2)}_t + S^{(2)}_y) - S^{(2)}_{xx}+ L S^{(2)}=0,\\
\label{nyo-03} &&L_t=2(\sigma_1|S^{(1)}|^2+\sigma_2|S^{(2)}|^2)_x,
\end{eqnarray}
where $\sigma_1=\pm1$, $\sigma_2=\pm1$, $S^{(1)},S^{(2)}\in \mathbf{C}$,
$L, t, x, y\in \mathbf{R}$,
was derived as a two-component generalization of the
2D YO system (the 2D long wave-short wave resonance interaction system)
by virtue of the reductive perturbation
method\cite{grimshaw1977modulation,oikawa1989two}.
The 2D coupled YO system can be written in the vector form:
\begin{eqnarray}
&&\textmd{i}(\mathbf{S}_t + \mathbf{S}_y) - \mathbf{S}_{xx}+ L \mathbf{S}=0,\\
&&L_t=2(\|\mathbf{S}\|^2)_x,
\end{eqnarray}
where $\mathbf{S}=(S^{(1)},S^{(2)})^{\mathrm T}$ and $\|\quad \|$ is defined as
\begin{equation}
\|\mathbf{S}\|^2=\sigma_1 S^{(1)}{S^{(1)}}^*+\sigma_2 S^{(2)}{S^{(2)}}^*\,.
\end{equation}
The above 2D coupled YO system can be generalized into a multi-component system, which is cast into the following vector form
\begin{eqnarray}
&&\textmd{i}(\mathbf{S}_t + \mathbf{S}_y) - \mathbf{S}_{xx}+ L \mathbf{S}=0,\\
&&L_t=2(\|\mathbf{S}\|^2)_x,
\end{eqnarray}
where $\mathbf{S}=(S^{(1)},S^{(2)},\cdots,S^{(M)})^{\mathrm{T}}$,
$\sigma_k=\pm1$ for $k=1,2,\cdots,M$,
and $\|\quad \|$ is defined as
\begin{equation}
\|\mathbf{S}\|^2=\sum_{k=1}^M
\sigma_k S^{(k)}{S^{(k)}}^*\,.
\end{equation}
The one-dimensional (1D) YO system was
proposed as a model equation for the interaction of
a Langmuir wave with an ion-acoustic wave in a plasma by Yajima and
Oikawa\cite{yajima1976formation},
which was also derived from several other physical
contexts\cite{grimshaw1977modulation,benney1976,djordjevic1977two,chowdhury2008long}.
The 1D YO system was solved exactly by the inverse scattering
transform method\cite{yajima1976formation,ma1979some} and the
(classical) Hirota's bilinear
method (which uses the perturbation expansion)\cite{ma1978complete,hirota2004direct}.
It admits both bright and dark soliton solutions.
The 2D YO system for the resonant interaction between a long surface wave
and a short internal wave in a two-layer fluid was presented and
the bright and dark soltion solutions are provided
by using the Hirota's bilinear method\cite{grimshaw1977modulation,oikawa1989two}.
The Painlev\'{e} property for the 2D YO system was investigated\cite{radha2005periodic} and
some special solutions such as positons, dromions, instantons and
periodic wave solutions
were constructed\cite{radha2005periodic,lai1999wave}.
For the 2D coupled case, the multi-bright
soliton solutions expressed by the Wronskian
to Eqs.(\ref{nyo-01})-(\ref{nyo-03}) were provided\cite{ohta2007two}. Later,
the bright N-soliton solutions in the Gram type determinant for the
multi-component YO system were obtained\cite{kanna2009higher,kanna2014general}.
Similar to the single component case, the Painlev\'{e} property and dromion
solutions
to Eqs.(\ref{nyo-01})-(\ref{nyo-03}) were discussed\cite{radha2009collision}.
In a recent paper by Kanna, Vijayajayanthi and Lakshmanan, one and two mixed soliton
solutions for the multi-component YO system were constructed\cite{kanna2012mixed}.
Very recently, the rogue wave solutions for the single YO system in 1D case were derived\cite{chow2013rogue,chen2014dark}.

However, to the best of our knowledge,
general multi-dark soliton solutions
for the multi-component 2D and 1D YO system have not been reported yet.
Moreover, general multi-dark soliton solutions for the multi-component 2D
soliton systems have never been previously reported in the literature.
In this paper, by using the reduction method of the KP hierarchy,
we derive and prove the general $N$-dark-dark soliton solutions to
Eqs.(\ref{nyo-01})-(\ref{nyo-03}) and their dynamics are discussed in detail.
Based on the KP theory, the general $N$-dark-dark soliton solutions
expressed by
either the Gram type or Wronski type determinant are obtained directly from
the $\tau$-functions of the KP hierarchy by means of reductions.
Similar to the 1D coupled nonlinear Schr\"odinger (NLS) equation
case \cite{ohta2011general},
it is very difficult to obtain multi-soliton solutions for the 1D
coupled YO system since
some non-trivial constraints for parameters need to be satisfied.
In this paper, we show that the general $N$-dark-dark soliton solutions for
the 1D coupled YO system can be obtained from the ones for the 2D
coupled YO system
by the reduction technique.
Thus the constraint condition is naturally obtained.

Kanna, Vijayajayanthi and Lakshmanan analyzed the bound states of the bright-dark
solitons\cite{kanna2012mixed}
and the bound states of the bright-bright solitons\cite{sakkaravarthi2013dynamics} for the 2D coupled YO system.
In this paper, we investigate the bound states of dark-dark solitons for
the 2D coupled YO system.
The bound states of dark-dark solitons for the 1D coupled NLS equation
with mixed focusing and defocusing nonlinearities was reported for the
first time by Ohta, Wang and Yang\cite{ohta2011general}.
The authors pointed out that the bound states of three or higher-dark-dark
solitons do not exist.
In the present paper, we show that the bound states of dark-dark
solitons can be
formed in the 2D coupled YO system including the stationary ones and moving ones.
For the stationary case, the bound states of arbitrary order dark-dark
solitons exist,
whereas, for the moving case, only the bound states of two dark-dark
solitons occur when the coefficients of nonlinear term take opposite signs.

The rest of the paper is organized as follows.
In Sect. 2, we briefly present the bilinearization procedure
for the 2D coupled YO system. The $N$-dark-dark soliton solutions
with the implicit dispersion relation are derived through the classical Hirota's
bilinear method which uses the perturbation expansion.
In Sect. 3, the general $N$-dark-dark soliton solutions
expressed by the Gram type and Wronski type determinants are obtained directly through
the reduction method of the KP hierarchy. Moreover, the general
$N$-dark-dark soliton solutions for the 1D coupled YO system are further
obtained by imposing a constraint on parameters.
Sect. 4 is devoted to the analysis of dynamics of one and two dark soltions,
which suggests that the energy of solitons is
completely transmitted through each component when two dark-dark solitons collide.
In Sect. 5, the bound states including
the stationary case and the moving case are discussed in detail.
In Sect. 6, the general $N$-dark-dark soliton solutions for
the 1D and 2D multi-component YO system are briefly presented.
Appendix A and B present the proofs of Lemma 2.1
and Lemma 2.4, respectively.

\section{Dark-dark soliton solutions of the two-dimensional coupled YO system }

Under the dependent variable transformation
\begin{equation}\label{nyo-04}
  S^{(1)}= \frac{G}{F},\ \ S^{(2)}= \frac{H}{F},\ \ L= -2(\log F)_{xx},
\end{equation}
Eqs. (\ref{nyo-01})--(\ref{nyo-03}) can be converted into the following bilinear equations:
\begin{eqnarray}
\label{nyo-05} \hspace{-0.5cm}&& [\textrm{i}(D_t+D_y)-D^2_x]G \cdot F=0,\\
\label{nyo-06} \hspace{-0.5cm}&& [\textmd{i}(D_t+D_y)-D^2_x]H \cdot F=0,\\
\label{nyo-07} \hspace{-0.5cm}&& (D_tD_x-2C) F \cdot F+2\sigma_1GG^*+2\sigma_2HH^*=0,
\end{eqnarray}
where  $G$ and $H$ are complex functions and $F$ is a real function,
$C$ is an arbitrary constant and ${}^*$ denotes the complex conjugation
hereafter.
The Hirota's $D$-operators are defined as
\[
D^n_x f(x)\cdot g(x)\equiv \bigg(\frac{\partial}{\partial x} - \frac{\partial}{\partial x'} \bigg)^n
f(x)g(x')\bigg|_{x=x'}.
\]

\subsection{Dark-dark soliton solutions by Hirota's direct method}

In this subsection, we look for soliton solutions by the Hirota's bilinear
method which uses the perturbation expansion\cite{hirota2004direct}. To this end,
we expand $G$, $H$ and $F$ in terms of power series of a small parameter $\epsilon$
\begin{eqnarray}
\label{nyo-08} \hspace{-0.5cm}&&G=\rho_1\exp(\textmd{i}\zeta_1)[1+\epsilon g_1+\epsilon^2 g_2 + \epsilon^3 g_3 + \cdots ]\,,\\
\label{nyo-09} \hspace{-0.5cm}&&H=\rho_2\exp(\textmd{i}\zeta_2)[1+\epsilon h_1+\epsilon^2 h_2 + \epsilon^3 h_3 + \cdots ]\,,\\
\label{nyo-10} \hspace{-0.5cm}&&F=1+\epsilon f_1+\epsilon^2 f_2 + \epsilon^3 f_3 +
 \cdots \,,
\end{eqnarray}
where $\zeta_j=\alpha_jx+\beta_jy-\delta_jt+\zeta_{j0}$ and $\rho_j,\alpha_j,\beta_j,\delta_j,\zeta_{j0}$, $(j{=}1,2)$ are real parameters.

Substituting (\ref{nyo-08})--(\ref{nyo-10}) into
(\ref{nyo-05})--(\ref{nyo-07}),
we obtain the following constraint condition:
\begin{equation}\label{nyo-11}
 C=\sigma_1\rho^2_1+\sigma_2\rho^2_2\,,  \ \  \delta_j=\beta_j-\alpha^2_j\,,\ \ j=1,2.
\end{equation}
Arranging each order of $\epsilon$ and solving the resultant set of
linear partial differential equations recursively,
we obtain the one-soliton solution:
\begin{eqnarray}
\label{nyo-12} \hspace{-0.5cm}&& G=\rho_1\exp(\textmd{i}\zeta_1)[1+ A \exp(\eta)]\,,\\
\label{nyo-13} \hspace{-0.5cm}&& H=\rho_2\exp(\textmd{i}\zeta_2)[1+ B \exp(\eta)]\,,\\
\label{nyo-14} \hspace{-0.5cm}&& F=1+  \exp(\eta),\ \ \eta=k_x x+k_y y+\omega t+\eta_0\,,
\end{eqnarray}
with
\begin{eqnarray*}
&&A=\frac{(2\alpha_1k_x-k_y-\omega)\textmd{i}-k_x^2}{(2\alpha_1k_x-k_y-\omega)\textmd{i}+k_x^2}\,,\\
&&B=\frac{(2\alpha_2k_x-k_y-\omega)\textmd{i}-k_x^2}{(2\alpha_2k_x-k_y-\omega)\textmd{i}+k_x^2}\,, \\
&& \omega k_x=\sigma_1\rho^2_1[2-(A+A^*)] + \sigma_2\rho^2_2[2-(B+B^*)]\,,
\end{eqnarray*}
where $k_x,k_y,\omega$ and $\eta_0$ are arbitrary complex constants.

Furthermore, we obtain the 2-soliton solution
\begin{eqnarray}
 \label{nyo-15}\hspace{-0.8cm}
&& G=\rho_1\exp(\textmd{i}\zeta_1)[1+ A_1 \exp(\eta_1) + A_2
\exp(\eta_2)\nonumber \\
\hspace{-0.8cm}
&&\qquad + C_{12}A_1A_2 \exp(\eta_1+\eta_2)]\,,\\
\label{nyo-16}\hspace{-0.8cm}&& H=\rho_2\exp(\textmd{i}\zeta_2)[1+ B_1
 \exp(\eta_1) + B_2 \exp(\eta_2) \nonumber\\
\hspace{-0.8cm}
&&\qquad + C_{12}B_1B_2 \exp(\eta_1+\eta_2)]\,,\\
\label{nyo-17}\hspace{-0.8cm}&& F=1+  \exp(\eta_1)+\exp(\eta_2)+C_{12}\exp(\eta_1+\eta_2)\,,
\end{eqnarray}
with
\begin{eqnarray*}
 &&A_j=\frac{(2\alpha_1k_{x,j}-k_{y,j}-\omega_j)\textmd{i}-k_{x,j}^2}{(2\alpha_1k_{x,j}-k_{y,j}-\omega_j)\textmd{i}+k_{x,j}^2}\,,\\
&&B_j=\frac{(2\alpha_2k_{x,j}-k_{y,j}-\omega_j)\textmd{i}-k_{x,j}^2}{(2\alpha_2k_{x,j}-k_{y,j}-\omega_j)\textmd{i}+k_{x,j}^2}\,,\\
 && \omega_j k_{x,j}=\rho^2_1[2-(A_j+A^*_j)] +
  \rho^2_2[2-(B_j+B^*_j)]\,, \\
&&\eta_j=k_{x,j}x+k_{y,j}y+\omega_j t+\eta_{j0}\,,
\end{eqnarray*}
and
\begin{eqnarray*}
\hspace{-0.8cm}&& C_{12}=-\frac{C_{12}^-}{C_{12}^{+}}\,,\\
\hspace{-0.8cm}&& C_{12}^{-}=(\theta_1-\theta_2)(\omega_1-\omega_2)+\sigma_1\rho^2_1(A_1A^*_2+A_2A^*_1-2)\\
\hspace{-0.8cm}&&\qquad +\sigma_2\rho^2_2(B_1B^*_2+B_2B^*_1-2)\,,\\
\hspace{-0.8cm}&&C_{12}^{+}=(\theta_1+\theta_2)(\omega_1+\omega_2)+\sigma_1\rho^2_1(A_1A_2+A^*_1A^*_2-2)\\
\hspace{-0.8cm}&&\qquad +\sigma_2\rho^2_2(B_1B_2+B^*_1B^*_2-2)\,,
\end{eqnarray*}
where $k_{x,j},k_{y,j},\omega_j$ and $\eta_{j0}$ ($j=1, 2$) are arbitrary complex constants.

In general, one can get the $N$-dark-dark soliton solutions of the 2D
coupled YO system
(\ref{nyo-01})--(\ref{nyo-03}):
\begin{eqnarray}
\label{nyo-18}\hspace{-0.8cm}&&G=\rho_1\textmd{e}^{\textmd{i}\zeta_1}\nonumber \\
\hspace{-0.8cm}&&\quad \times \sum_{\mu=0,1} \exp\left(\sum^N_{j=1} \mu_j (\eta_j+a_j) + \sum^N_{1\leq j<l}\mu_j\mu_lc_{jl}\right),\\
\label{nyo-19}\hspace{-0.8cm}&&H=\rho_2\textmd{e}^{\textmd{i}\zeta_2}\nonumber\\
\hspace{-0.8cm}&& \quad \times \sum_{\mu=0,1}  \exp\left(\sum^N_{j=1} \mu_j (\eta_j+b_j) + \sum^N_{1\leq j<l}\mu_j\mu_lc_{jl}\right),\\
\label{nyo-20}\hspace{-0.8cm}&&F=\sum_{\mu=0,1}  \exp\left(\sum^N_{j=1} \mu_j \eta_j + \sum^N_{1\leq j<l}\mu_j\mu_lc_{jl}\right),
\end{eqnarray}
with
\begin{eqnarray*}
\hspace{-0.5cm} && \sigma_1\rho^2_1+\sigma_2\rho^2_2=C\,,\\
\hspace{-0.5cm} && \zeta_k=\alpha_kx+\beta_ky-(\beta_k-\alpha^2_k)t+\zeta_{k0}\,, \ \ k{=}1,2\\
\hspace{-0.5cm} &&\textmd{e}^{a_j}\equiv A_j=\frac{(2\alpha_1k_{x,j}-k_{y,j}-\omega_j)\textmd{i}-k_{x,j}^2}{(2\alpha_1k_{x,j}-k_{y,j}-\omega_j)\textmd{i}+k_{x,j}^2}\,,\\
\hspace{-0.5cm}&&\textmd{e}^{b_j}\equiv B_j=\frac{(2\alpha_2k_{x,j}-k_{y,j}-\omega_j)\textmd{i}-k_{x,j}^2}{(2\alpha_2k_{x,j}-k_{y,j}-\omega_j)\textmd{i}+k_{x,j}^2} \,,\\
\hspace{-0.5cm} && \omega_j k_{x,j}=\sigma_1\rho^2_1[2-(A_j+A^*_j)] +\sigma_2 \rho^2_2[2-(B_j+B^*_j)]\,,
\end{eqnarray*}
and
\begin{eqnarray*}
\hspace{-0.8cm}&&C_{jl} \equiv \textmd{e} ^{c_{jl}}=-\frac{C_{jl}^-}{C_{jl}^+}\,,\\
\hspace{-0.8cm}&&C_{jl}^-=(k_{x,j}-\theta_l)(\omega_j-\omega_l)+\sigma_1\rho^2_1(A_jA^*_l+A_lA^*_j-2)\\
\hspace{-0.8cm}&&\qquad +\sigma_2\rho^2_2(B_jB^*_l+B_lB^*_j-2)\,,\\
\hspace{-0.8cm}&&C_{jl}^+=(k_{x,j}+\theta_l)(\omega_j+\omega_l)+\sigma_1\rho^2_1(A_jA_l+A^*_jA^*_l-2)\\
\hspace{-0.8cm}&&\qquad +\sigma_2\rho^2_2(B_jB_l+B^*_jB^*_l-2)\,,
\end{eqnarray*}
where the notation $\sum_{\mu=0,1}$ represents all possible combinations
$\mu_j=0,1$ and $\eta_j=k_{x,j}x+k_{y,j}y+\omega_j t+\eta_{j0}$ for
$j=1,2,3\ldots N$.

\subsection{General $N$-dark-dark soliton solutions in the Gram determinant form}

In this subsection, we construct an alternative form of the $N$-soliton solution based on the KP hierarchy reduction method.

\begin{lemma}
\label{lemma1}
The following bilinear equations in the KP hierarchy\cite{ohta2011general}:
\begin{eqnarray}
\label{nyo-21}\hspace{-0.8cm}&& (D_{x_2}-D^2_{x_1}-2a D_{x_1})\tau(k+1,l)\cdot \tau(k,l)=0\,,\\
\hspace{-0.8cm}&& \left(\frac{1}{2}D_{x_1}D_{x_{-1}}-1\right)
\tau(k,l)\cdot \tau(k,l)\nonumber\\
\label{nyo-22}\hspace{-0.8cm}&&\qquad \qquad =-\tau(k+1,l)\tau(k-1,l)\,,\\
\label{nyo-23}\hspace{-0.8cm}&& (D_{x_2}-D^2_{x_1}-2b D_{x_1})\tau(k,l+1)\cdot \tau(k,l)=0\,,\\
\hspace{-0.8cm}&& \left(\frac{1}{2}D_{x_1}D_{y_{-1}}-1\right)
\tau(k,l)\cdot \tau(k,l)\nonumber\\
\label{nyo-24}\hspace{-0.8cm}&&\qquad \qquad =-\tau(k,l+1)\tau(k,l-1)\,,
\end{eqnarray}
where $a$ and $b$ are complex constants, and $k$ and $l$ are integers,
have the Gram type determinant solutions
\begin{eqnarray}
\label{nyo-25} \tau(k,l)=\left|
\begin{array}{c} m_{ij}(k,l) \end{array}
\right| _{1\leq i,j\leq N}\,,
\end{eqnarray}
where the entries of the determinant are given by
\begin{eqnarray*}
&& m_{ij}(k,l)=c_{ij}+\int \phi_i(k,l)\psi_j(k,l)\ dx_1,\\
&& \phi_i(k,l)=(p_i-a)^k(p_i-b)^l\exp(\theta_i),\\
&& \psi_i(k,l)=\left(-\frac{1}{q_i+a}\right)^k \left(-\frac{1}{q_i+b}\right)^l\exp(\tilde{\theta}_i)\,,
\end{eqnarray*}
with
\begin{eqnarray*}
&& \theta_i=\frac{1}{p_i-a}x_{-1} + \frac{1}{p_i-b}y_{-1} + p_ix_1 + p^2_ix_2 + \theta_{i0},\\
&& \tilde{\theta}_j=\frac{1}{q_i+a}x_{-1} + \frac{1}{q_i+b}y_{-1} + q_ix_1 - q^2_ix_2 + \tilde{\theta}_{i0},
\end{eqnarray*}
where $c_{ij}, p_i, q_i, \theta_{i0}$ and $\tilde{\theta}_{i0}$
 ($i,j=1,2,\cdots, N$) are
arbitrary complex constants.
\end{lemma}
The proof is given in the Appendix.

Now we consider the reduction of the above bilinear equations in order to
derive the general dark soliton solution.
Assuming $x_{-1}$, $y_{-1}$, $x_1$ are real, $x_2$, $a(=\textmd{i}\alpha_1)$, $b(=\textmd{i}\alpha_2)$ are pure imaginary and $q_i=p^*_i$, $\tilde{\theta}_{j0}=\theta^*_{j0}$, $c_{ij}=c^*_{ij}=\delta_{ij}$, one can get
\begin{eqnarray}
&&\tilde{\theta}_j=\theta^*_j,\ \ m_{ji}(k,l)=m^*_{ij}(-k,-l),\nonumber\\
&&\tau(k,l)=\tau^*(-k,-l).\label{nyo-26}
\end{eqnarray}
Hence,
Eqs.(\ref{nyo-21})-(\ref{nyo-24}) can be recast into
\begin{eqnarray}
\label{nyo-27} && (D_{x_2}-D^2_{x_1}-2\textmd{i}\alpha_1D_{x_1})g\cdot f=0,\\
\label{nyo-28} && \left(\frac{1}{2}D_{x_1}D_{x_{-1}}-1\right)f\cdot f=-gg^*,\\
\label{nyo-29} && (D_{x_2}-D^2_{x_1}-2\textmd{i}\alpha_2D_{x_1})h\cdot f=0,\\
\label{nyo-30} && \left(\frac{1}{2}D_{x_1}D_{y_{-1}}-1\right)f\cdot f=-hh^*,
\end{eqnarray}
by defining
\begin{eqnarray}
&&  f=\tau(0,0),\ \ g=\tau(1,0),\ \ h=\tau(0,1),\nonumber \\
\label{nyo-31} &&g^*=\tau(-1,0),\ \ h^*=\tau(0,-1)\,.
\end{eqnarray}

By introducing the independent variable transformation
\begin{eqnarray}
  &&  x_1=x,\ \ x_2=-\textmd{i}y,\nonumber\\
\label{nyo-33}  && x_{-1}=\sigma_1{\rho^2_1}(t-y),\ \ y_{-1}=\sigma_2{\rho^2_2}(t-y),
\end{eqnarray}
i.e.,
\begin{eqnarray}
&&\partial_x=\partial_{x_1},\ \ \partial_y= -\textmd{i}\partial_{x_2}
  -\sigma_1{\rho^2_1}\partial_{x_{-1}}-
  \sigma_2{\rho^2_2}\partial_{y_{-1}},\nonumber\\
\label{nyo-34}&& \partial_t=\sigma_1{\rho^2_1}\partial_{x_{-1}} +\sigma_2{\rho^2_2} \partial_{y_{-1}},
\end{eqnarray}
Eqs.(\ref{nyo-27})-(\ref{nyo-30}) become
\begin{eqnarray}
 \label{nyo-35}\hspace{-0.8cm}&& [\textmd{i}(D_t+D_y-2\alpha_1D_x)-D^2_x]g \cdot f=0\,,\\
 \label{nyo-35-2}\hspace{-0.8cm}&&[\textmd{i}(D_t+D_y-2\alpha_1D_x)+D^2_x]g^* \cdot f=0\,,\\
 \label{nyo-36}\hspace{-0.8cm}&& [\textmd{i}(D_t+D_y-2\alpha_2D_x)-D^2_x]h \cdot f=0\,,\\
 \label{nyo-36-2}\hspace{-0.8cm}&&[\textmd{i}(D_t+D_y-2\alpha_2D_x)+D^2_x]h^* \cdot f=0\,,\\
\hspace{-0.8cm}&&
  [D_tD_x-2(\sigma_1\rho^2_1+\sigma_2\rho^2_2)]f \cdot f\nonumber\\
 \label{nyo-37}\hspace{-0.8cm}&&\qquad +2\sigma_1\rho^2_1gg^*+2\sigma_2\rho^2_2hh^*=0\,.
\end{eqnarray}
By virtue of the following dependent variable transformation
\begin{equation}\label{nyo-38}
    S^{(1)}= \rho_1\textmd{e}^{\textmd{i}\zeta_1}\frac{g}{f}\,,\ \ S^{(2)}= \rho_2\textmd{e}^{\textmd{i}\zeta_2}\frac{h}{f}\,,\ \
  L= -2\left(\log f\right)_{xx}\,,
\end{equation}
with $\zeta_j=\alpha_jx+\beta_jy-(\beta_j-\alpha^2_j)t+\zeta_{j0}$ for
$j{=}1,2$,
the bilinear equations (\ref{nyo-35})--(\ref{nyo-37}) are then
transformed into the 2D
YO system (\ref{nyo-01})-(\ref{nyo-03}).
Hence we immediately have the following theorem for the general
$N$-dark-dark
soliton solutions of Eqs. (\ref{nyo-01})-(\ref{nyo-03}).

\begin{theorem}
The $N$-dark-dark soliton solutions for the 2D coupled YO system (\ref{nyo-01})-(\ref{nyo-03}) are
\begin{eqnarray}
\label{nyo-39}&& S^{(1)}= \rho_1\textmd{e}^{\textmd{i}[\alpha_1x+\beta_1y-(\beta_1-\alpha^2_1)t+\zeta_{10}]}\frac{g}{f}\,,\\
\label{nyo-40}&& S^{(2)}= \rho_2\textmd{e}^{\textmd{i}[\alpha_2x+\beta_2y-(\beta_2-\alpha^2_2)t+\zeta_{20}]}\frac{h}{f}\,,\\
\label{nyo-41}&& L= -2\left(\log f\right)_{xx}\,,
\end{eqnarray}
where $f,g$ and $h$ are Gram determinants given by
\begin{eqnarray*}
&&f=\Bigg|  \delta_{ij} + \frac{1}{p_i+p^*_j} \textmd{e}^{\xi_i+\xi^*_j}  \Bigg|_{N\times N}, \\ &&
  g=\Bigg| \delta_{ij} + \left(-\frac{p_i-\textmd{i}\alpha_1}{p^*_j+\textmd{i}\alpha_1}\right)  \frac{1}{p_i+p^*_j} \textmd{e}^{\xi_i+\xi^*_j} \Bigg|_{N\times N},\ \ \\
&& h=\Bigg| \delta_{ij} + \left(-\frac{p_i-\textmd{i}\alpha_2}{p^*_j+\textmd{i}\alpha_2}\right)  \frac{1}{p_i+p^*_j} \textmd{e}^{\xi_i+\xi^*_j}  \Bigg|_{N\times N},\ \
\end{eqnarray*}
with
\begin{eqnarray*}
\hspace{-0.5cm}&& \xi_j=p_jx -
 \Big(\frac{\sigma_1\rho^2_1}{p_j-\textmd{i}\alpha_1}+\frac{\sigma_2\rho^2_2}{p_j-\textmd{i}\alpha_2}
 +\textmd{i}p^2_j\Big)y \\
\hspace{-0.5cm}&&
\qquad \qquad \qquad + \Big(\frac{\sigma_1\rho^2_1}{p_j-\textmd{i}\alpha_1}+\frac{\sigma_2\rho^2_2}{p_j-\textmd{i}\alpha_2} \Big)t  + \xi_{j0},\ \ \  \
\end{eqnarray*}
where $\alpha_k,\beta_k, \rho_k,\zeta_{k0}$ ($k{=}1,2$) are real
constants, and $p_j,\xi_{j0}$ are arbitrary complex constants.
\end{theorem}

\begin{remark}
By putting the parameters in
(\ref{nyo-18})-(\ref{nyo-20}) as
\begin{eqnarray*}
 \hspace{-0.8cm}&&k_{x,j}=p_j+p^*_j, \quad k_{y,j}=-\omega_j-\textmd{i}(p^2_j-p^{*2}_j),\\
 \hspace{-0.8cm}&&\omega_j=
  \frac{\sigma_1\rho^2_1(p_j+p^*_j)}{|p_j-\textmd{i}\alpha_1|^2}  +
  \frac{\sigma_2\rho^2_2(p_j+p^*_j)}{|p_j-\textmd{i}\alpha_2|^2},\\
\hspace{-0.8cm}&&\eta_{j0}=\xi_{j0}+\xi^*_{j0}+\frac{1}{p_j+p^*_j},
\end{eqnarray*}
the $N$-dark-dark soliton solutions (\ref{nyo-18})-(\ref{nyo-20}) are
equivalent to the Gram type
determinant solutions (\ref{nyo-39})-(\ref{nyo-41}).
\end{remark}

\subsection{General $N$-dark-dark soliton solutions in the Wronskian form}

In this subsection, we show that the general $N$-dark-dark soliton
solutions
for the 2D coupled YO system (\ref{nyo-01})-(\ref{nyo-03}) can be
expressed in the Wronskian from.

\begin{lemma}
\label{lemma2}
The following Wronskian satisfies
the bilinear equations of the KP hierarchy  (\ref{nyo-21})-(\ref{nyo-24}):
\begin{eqnarray}
\hspace{-0.8cm}&&\tau(k,l)\nonumber \\
\hspace{-0.8cm}&& \quad =\left| \begin{array}{ccccc}
\varphi_1(k,l) & \partial_{x_1}\varphi_1(k,l) & \cdots  & \partial^{(N-1)}_{x_1}\varphi_1(k,l)  \\
\varphi_2(k,l) & \partial_{x_1}\varphi_2(k,l) & \cdots  & \partial^{(N-1)}_{x_1}\varphi_2(k,l)  \\
\vdots & \vdots & \cdots  & \vdots  \\
\varphi_N(k,l) & \partial_{x_1}\varphi_N(k,l) & \cdots  & \partial^{(N-1)}_{x_1}\varphi_N(k,l)
\end{array}
\right|,\nonumber\\
\hspace{-0.8cm}&&\label{nyo-42}
\end{eqnarray}
with
\begin{eqnarray*}
&& \varphi_i(k,l)=(p_i-a)^k(p_i-b)^l\exp(\theta_i)\\
&& \qquad \qquad \qquad + (q_i-a)^k (q_i-b)^l\exp(\tilde{\theta}_i),\\
&& \theta_i=\frac{1}{p_i-a}x_{-1} + \frac{1}{p_i-b}y_{-1} + p_ix_1 + p^2_ix_2 + \theta_{i0},\\
&& \tilde{\theta}_i=\frac{1}{q_i-a}x_{-1} + \frac{1}{q_i-b}y_{-1} + q_ix_1 + q^2_ix_2 + \tilde{\theta}_{i0},
\end{eqnarray*}
where $p_j, q_j, \theta_{i0}$ and $\tilde{\theta}_{i0}$ are arbitrary complex constants.
\end{lemma}
We provide the proof in the Appendix.

Next, we proceed to reductions.
By applying the complex conjugate conditions
\begin{eqnarray}\label{nyo-43}
p^*_i=-q_i,\ \ \theta^*_{i0}=-\tilde{\theta}'_{i0},
\end{eqnarray}
with $\exp(\tilde{\theta}_{i0}) =\left( \prod^{N}_{k=1, k\neq i} \frac{p_i-q_k}{q_i-q_k}   \right) \exp(\tilde{\theta}'_{i0})$ and the determinant formula\cite{marunoohta2006}
\begin{eqnarray}\label{nyo-44}
\nonumber \hspace{-0.8cm}  && \det_{1\leq i,j \leq N} \left(  p^{j-1}_i \mathcal{A}_i + q^{j-1}_i \mathcal{B}_i  \right) \\
\hspace{-0.8cm}&& =\left( \Delta(q_1,q_2,\cdots, q_N)
 \prod^N_{i=1}\mathcal{B}_i \right) \nonumber \\
\hspace{-0.8cm}&&\times \sum^{N}_{M=0} \sum_{1\leq i_1 <
 \cdots < i_M \leq N} \left( \prod_{1\leq\mu< \nu\leq M}  \frac{(p_{i_{\mu}} {-} p_{i_{\nu}})(q_{i_{\mu}} {-} q_{i_{\nu}})}{(p_{i_{\mu}} {-} q_{i_{\nu}})(q_{i_{\mu}} {-} p_{i_{\nu}})} \right)\nonumber\\
\hspace{-0.8cm}&&\times\prod^M_{\nu =1}\frac{\mathcal{A}_{i_{\nu}}}{\mathcal{B}'_{i_{\nu}}},\ \ \ \ \ \ \
\end{eqnarray}
where $ \mathcal{B}_i= \left( \prod^{N}_{k=1, k\neq i} \frac{p_i-q_k}{q_i-q_k}   \right) \mathcal{B}'_i$ and
$\Delta$ is the Vandermonde determinant,
the following relation can be derived
\begin{eqnarray}
\hspace{-0.8cm}&&\tau(0,0) =\mathcal{G} \sum^{N}_{M=0} \sum_{1\leq i_1  <
 \cdots < i_M \leq N}
\left| \mathcal{F}(i_{\mu},i_{\nu}) \right|^2\nonumber\\
\label{nyo-45}\hspace{-0.8cm}&&\quad \times
\prod^M_{\nu =1}\exp(\theta_{i_{\nu}}+\theta^*_{i_{\nu}}),\\
\hspace{-0.8cm}&&\tau(1,0) = \mathcal{C}_a \cdot \mathcal{G} \sum^{N}_{M=0} \sum_{1\leq i_1  < \cdots < i_M \leq N}
\left| \mathcal{F}(i_{\mu},i_{\nu}) \right|^2\nonumber\\
\label{nyo-46}\hspace{-0.8cm}&&\quad \times\prod^M_{\nu =1} \left( {-} \frac{p_{i_{\nu}}-a}{p^*_{i_{\nu}}+a} \right) \exp(\theta_{i_{\nu}}+\theta^*_{i_{\nu}}),\\
\hspace{-0.8cm}&&\tau(-1,0) =\frac{1}{\mathcal{C}_a} \cdot  \mathcal{G} \sum^{N}_{M=0} \sum_{1\leq i_1  < \cdots < i_M \leq N} \left| \mathcal{F}(i_{\mu},i_{\nu}) \right|^2\nonumber\\
\label{nyo-47}\hspace{-0.8cm}&&\quad \times\prod^M_{\nu =1} \left( {-} \frac{p^*_{i_{\nu}}+a}{p_{i_{\nu}}-a} \right) \exp(\theta_{i_{\nu}}+\theta^*_{i_{\nu}}),\ \ \ \ \ \ \ \\
\hspace{-0.8cm}&&\tau(0,1) = \mathcal{C}_b \cdot \mathcal{G} \sum^{N}_{M=0} \sum_{1\leq i_1  < \cdots < i_M \leq N}
\left| \mathcal{F}(i_{\mu},i_{\nu}) \right|^2\nonumber\\
\label{nyo-48}\hspace{-0.8cm}&&\quad \times\prod^M_{\nu =1} \left( {-} \frac{p_{i_{\nu}}-b}{p^*_{i_{\nu}}+b}\right) \exp(\theta_{i_{\nu}}+\theta^*_{i_{\nu}}),\\
\hspace{-0.8cm}&&\tau(0,-1) =\frac{1}{\mathcal{C}_b} \cdot  \mathcal{G} \sum^{N}_{M=0}
 \sum_{1\leq i_1  < \cdots < i_M \leq N} \left|
					  \mathcal{F}(i_{\mu},i_{\nu})
					 \right|^2\nonumber\\
\label{nyo-49}\hspace{-0.8cm}&&\quad \times
\prod^M_{\nu =1} \left( {-} \frac{p^*_{i_{\nu}}+b}{p_{i_{\nu}}-b}\right) \exp(\theta_{i_{\nu}}+\theta^*_{i_{\nu}}),\ \
\end{eqnarray}
where
\begin{eqnarray*}
 &&\mathcal{G}=\Delta(-p^*_1,-p^*_2,\cdots, -p^*_N) \prod^N_{i=1} \left(
								   \prod^{N}_{k=1,
								   k\neq
								   i}
								   -\frac{p_i+p^*_k}{p^*_i-p^*_k}
								  \right)\\
&& \qquad \qquad \times  \exp(-\theta^*_i),\\
 &&\mathcal{C}_a=\prod^N_{i=1}(-1)^N(p^*_i+a),\ \
  \mathcal{C}_b=\prod^N_{i=1}(-1)^N(p^*_i+b), \\
 &&\mathcal{F}(i_{\mu},i_{\nu})= \prod_{1\leq\mu<\nu\leq M}  \frac{(p_{i_{\mu}} {-} p_{i_{\nu}})}{(p_{i_{\mu}} {+} p^*_{i_{\nu}})}.
\end{eqnarray*}


Next, setting
\begin{eqnarray}\label{nyo-50}
&& f=\frac{\tau(0,0)}{\mathcal{G}},\ \
 g=\frac{\tau(1,0)}{\mathcal{C}_a \mathcal{G}},\ \
 h=\frac{\tau(0,1)}{\mathcal{C}_b\mathcal{G}},\nonumber\\
&& g^*=\frac{\mathcal{C}_a \tau(-1,0)}{\mathcal{G}},\ \ h^*=\frac{\mathcal{C}_b \tau(0,-1)}{\mathcal{G}},
\end{eqnarray}
Eqs. (\ref{nyo-21})-(\ref{nyo-24}) become
\begin{eqnarray}
\label{nyo-51}&& (D_{x_2}-D^2_{x_1}-2a D_{x_1})g\cdot f=0,\\
\label{nyo-52}&& \left(\frac{1}{2}D_{x_1}D_{x_{-1}}-1\right)f\cdot f=-gg^*,\\
\label{nyo-53}&& (D_{x_2}-D^2_{x_1}-2b D_{x_1})h\cdot f=0,\\
\label{nyo-54}&& \left(\frac{1}{2}D_{x_1}D_{y_{-1}}-1\right)f\cdot f=-hh^*,
\end{eqnarray}
which are nothing but the bilinear equations (\ref{nyo-27})-(\ref{nyo-30}) if $a=\textmd{i}\alpha_1$, $b=\textmd{i}\alpha_2$. Then, by applying the same transformations of independent variables (\ref{nyo-33}) and dependent variables (\ref{nyo-38}), the 2D coupled YO system (\ref{nyo-01})-(\ref{nyo-03}) can be obtained.

In summary, we obtain an alternative form of the $N$-dark-dark soliton solution of Eqs. (\ref{nyo-01})-(\ref{nyo-03}) in the following theorem:

\begin{theorem}
The $N$-dark-dark soliton solutions for the 2D coupled YO system (\ref{nyo-01})-(\ref{nyo-03}) are
\begin{eqnarray}
\label{nyo-55}&& S^{(1)}= \rho_1\textmd{e}^{\textmd{i}[\alpha_1x+\beta_1y-(\beta_1-\alpha^2_1)t+\zeta_{10}]}\frac{g}{f},\\
\label{nyo-56}&& S^{(2)}= \rho_2\textmd{e}^{\textmd{i}[\alpha_2x+\beta_2y-(\beta_2-\alpha^2_2)t+\zeta_{20}]}\frac{h}{f},\\
\label{nyo-57}&& L= -2\left(\log f\right)_{xx},
\end{eqnarray}
where $f,g$ and $h$ are Wronskians expressed by the form
\begin{eqnarray*}
&&f=\frac{1}{\mathcal{G}}\left| \begin{array}{ccccc}
\varphi_1 & \partial_{x_1}\varphi_1 & \cdots  & \partial^{(N-1)}_{x_1}\varphi_1  \\
\varphi_2 & \partial_{x_1}\varphi_2 & \cdots  & \partial^{(N-1)}_{x_1}\varphi_2  \\
\vdots & \vdots & \cdots  & \vdots  \\
\varphi_N & \partial_{x_1}\varphi_N & \cdots  & \partial^{(N-1)}_{x_1}\varphi_N
\end{array} \right|, \\ &&
  g=\frac{1}{\mathcal{C}_1 \mathcal{G}}\left| \begin{array}{ccccc}
\bar{\varphi}_1 & \partial_{x_1}\bar{\varphi}_1 & \cdots  & \partial^{(N-1)}_{x_1}\bar{\varphi}_1  \\
\bar{\varphi}_2 & \partial_{x_1}\bar{\varphi}_2 & \cdots  & \partial^{(N-1)}_{x_1}\bar{\varphi}_2  \\
\vdots & \vdots & \cdots  & \vdots  \\
\bar{\varphi}_N & \partial_{x_1}\bar{\varphi}_N & \cdots  & \partial^{(N-1)}_{x_1}\bar{\varphi}_N
\end{array} \right|, \ \ \\
&& h=\frac{1}{\mathcal{C}_2 \mathcal{G}}\left| \begin{array}{ccccc}
\tilde{\varphi}_1 & \partial_{x_1}\tilde{\varphi}_1 & \cdots  & \partial^{(N-1)}_{x_1}\tilde{\varphi}_1  \\
\tilde{\varphi}_2 & \partial_{x_1}\tilde{\varphi}_2 & \cdots  & \partial^{(N-1)}_{x_1}\tilde{\varphi}_2  \\
\vdots & \vdots & \cdots  & \vdots  \\
\tilde{\varphi}_N & \partial_{x_1}\tilde{\varphi}_N & \cdots  & \partial^{(N-1)}_{x_1}\tilde{\varphi}_N
\end{array} \right|,\ \
\end{eqnarray*}
with
\begin{eqnarray*}
\hspace{-0.8cm}&&
\mathcal{G}=\Delta(-p^*_1,-p^*_2,\cdots, -p^*_N) \\
\hspace{-0.8cm}
&&\qquad \qquad  \times \prod^N_{j=1} \bigg( \prod^{N}_{k=1, k\neq j} -\frac{p_j+p^*_k}{p^*_j-p^*_k}   \bigg) \exp(-\xi^*_j),\\
\hspace{-0.8cm}&&
\mathcal{C}_1=\prod^N_{j=1}(-1)^N(p^*_j+\textmd{i}\alpha_1),\ \ \mathcal{C}_2=\prod^N_{j=1}(-1)^N(p^*_j+\textmd{i}\alpha_2),
\end{eqnarray*}
and
\begin{eqnarray*}
\hspace{-0.8cm}&& \varphi_j=\exp(\xi_j)+\exp(-\xi^*_j),\\
\hspace{-0.8cm}&& \bar{\varphi}_j=(p_j-\textmd{i}\alpha_1)\exp(\xi_j)-(p^*_j+\textmd{i}\alpha_1)\exp(-\xi^*_j),\\
\hspace{-0.8cm}&& \tilde{\varphi}_j=(p_j-\textmd{i}\alpha_2)\exp(\xi_i)-(p^*_j+\textmd{i}\alpha_2)\exp(-\xi^*_j),\\
\hspace{-0.8cm}&& \xi_j=p_jx -
 \Big(\frac{\sigma_2\rho^2_1}{p_j-\textmd{i}\alpha_1}+\frac{\sigma_2\rho^2_2}{p_j-\textmd{i}\alpha_2}
 +\textmd{i}p^2_j\Big)y \\
\hspace{-0.8cm}&& \qquad \qquad \quad + \Big(\frac{\sigma_1\rho^2_1}{p_j-\textmd{i}\alpha_1}+\frac{\sigma_2\rho^2_2}{p_j-\textmd{i}\alpha_2} \Big)t  + \xi_{j0},\ \ \ \
\end{eqnarray*}
where $\alpha_k,\beta_k, \rho_k,\zeta_{k0},(k{=}1,2)$ are real constants, and $p_j,\xi_{j0}$ are complex constants.
\end{theorem}

\section{General $N$-dark-dark soliton solutions of the one-dimensional coupled YO system}
The general $N$-dark-dark soliton solutions for the 1D coupled YO system
can be derived from the one for 2D coupled YO system by further
reductions. In what follows, we show the detailed process.

First, it is noted that the Gram determinant solution of the bilinear equations (\ref{nyo-21})-(\ref{nyo-24}) in the KP hierarchy can be rewritten as
\begin{eqnarray}\label{nyo-58}
\hspace{-0.8cm}\nonumber && \tau(k,l) \\
\hspace{-0.8cm}\nonumber&& = \Bigg| \delta_{ij} + \left(-\frac{p_i-\textmd{i}\alpha_1}{p^*_j+\textmd{i}\alpha_1}\right)^k \left(-\frac{p_i-\textmd{i}\alpha_2}{p^*_j+\textmd{i}\alpha_2}\right)^l  \frac{1}{p_i+p^*_j} \textmd{e}^{\xi_i+\xi^*_j} \Bigg|\\
\hspace{-0.8cm}&&= \exp\left(\sum_{k=1}^N (\xi_k+\xi^*_k)\right) \nonumber\\
\hspace{-0.8cm}&&\times \Bigg| \delta_{ij} \textmd{e}^{-\xi_i-\xi^*_i} +
 \left(-\frac{p_i-\textmd{i}\alpha_1}{p^*_j+\textmd{i}\alpha_1}\right)^k\left(-\frac{p_i-\textmd{i}\alpha_2}{p^*_j+\textmd{i}\alpha_2}\right)^l
 \frac{1}{p_i+p^*_j}  \Bigg|\,,\nonumber\\
\hspace{-0.8cm}&&
\end{eqnarray}
with
\begin{eqnarray*}
\hspace{-0.8cm} \nonumber && \xi_i+\xi^*_i=\bigg(\frac{1}{p_i-\textmd{i}\alpha_1} + \frac{1}{p^*_i+\textmd{i}\alpha_1}  \bigg) x_{-1}\\
\hspace{-0.8cm}\nonumber &&\qquad  + \bigg(\frac{1}{p_i-\textmd{i}\alpha_2} + \frac{1}{p^*_i+\textmd{i}\alpha_2} \bigg) y_{-1}\\
\hspace{-0.8cm}&&\qquad  + (p_i + p^*_i)x_1 + (p^2_i-p^{*2}_i )x_2 + \xi_{i0} +
 \xi^*_{i0}\,.
\end{eqnarray*}
Thus, if $p_i$ satisfies the constraint condition:
\begin{eqnarray}\label{nyo-59}
\hspace{-0.8cm}&&\sigma_1\rho^2_1 \bigg( \frac{1}{p_i-\textmd{i}\alpha_1}
 + \frac{1}{p^*_i+\textmd{i}\alpha_1}  \bigg) \nonumber\\
\hspace{-0.8cm}&&\, + \sigma_2\rho^2_2 \bigg(
 \frac{1}{p_i-\textmd{i}\alpha_2} + \frac{1}{p^*_i+\textmd{i}\alpha_2}
 \bigg) = - \textmd{i} (p^2_i-p^{*2}_i),
\end{eqnarray}
i.e.,
\begin{eqnarray}\label{nyo-60}
 \frac{\sigma_1\rho^2_1}{|p_i-\textmd{i}\alpha_1|^2}  +  \frac{\sigma_2\rho^2_2}{|p_i-\textmd{i}\alpha_2|^2}  = - \textmd{i} (p_i-p^{*}_i),
\end{eqnarray}
then we have
\begin{eqnarray}\label{nyo-61}
(\sigma_1\rho^2_1 \partial_{x_{-1}} + \sigma_2\rho^2_2 \partial_{y_{-1}} ) \tau(k,l) = -\textmd{i} \partial_{x_2} \tau(k,l)\,,
\end{eqnarray}
which implies
\begin{eqnarray}\label{nyo-62}
(\sigma_1\rho^2_1 \partial_{x_{-1}} + \sigma_2\rho^2_2 \partial_{y_{-1}} ) f = -\textmd{i} \partial_{x_2} f,
\end{eqnarray}
by using $f=\tau(0,0)$. Moreover, differentiating with respect to $x_1$ once, we have
\begin{eqnarray}\label{nyo-63}
\sigma_1\rho^2_1 f_{x_1x_{-1}} + \sigma_2\rho^2_2 f_{x_1y_{-1}}  = -\textmd{i} f_{x_1x_2}.
\end{eqnarray}
Notice that Eqs.(\ref{nyo-28}) and (\ref{nyo-30}) can be rewritten as
\begin{eqnarray}
\label{nyo-64} &&  f_{x_1x_{-1}}f - f_{x_1}f_{x_{-1}}- f^2=-gg^*,\\
\label{nyo-65} &&  f_{x_1y_{-1}}f - f_{x_1}f_{y_{-1}}- f^2=-hh^*\,.
\end{eqnarray}
From the above relations, we have
\begin{eqnarray}
&&-\textmd{i}(f_{x_1x_2}f-f_{x_2}f_{x_1})- (\sigma_1\rho^2_1+
 \sigma_2\rho^2_2)f^2\nonumber\\
&& \qquad +\sigma_1\rho^2_1gg^*+ \sigma_2\rho^2_2hh^*=0,\label{nyo-66}
\end{eqnarray}
or the bilinear form
\begin{eqnarray}
&&-\textmd{i}D_{x_1}D_{x_2}f\cdot f- 2(\sigma_1\rho^2_1+
 \sigma_2\rho^2_2)f^2\nonumber\\
&& \quad +2\sigma_1\rho^2_1gg^*+ 2\sigma_2\rho^2_2hh^*=0\,.\label{nyo-67}
\end{eqnarray}
Finally, by using the transformations of independent variables
\begin{eqnarray}\label{nyo-68}
x_1=x,\ \ x_2=-\textmd{i}t,
\end{eqnarray}
i.e.,
\begin{eqnarray}\label{nyo-69}
\partial_{x_1}=\partial_x,\ \ \partial_{x_2}=\textmd{i}\partial_t\,,
\end{eqnarray}
Eqs. (\ref{nyo-27})-(\ref{nyo-30}) become
\begin{eqnarray}
\label{nyo-70}\hspace{-0.8cm} && [\textmd{i}(D_t-2\alpha_1D_x)-D^2_x]g \cdot f=0\,,\\
\label{nyo-70-2}\hspace{-0.8cm}&&[\textmd{i}(D_t-2\alpha_1D_x)+D^2_x]g^* \cdot f=0\,,\\
\label{nyo-71}\hspace{-0.8cm} && [\textmd{i}(D_t-2\alpha_2D_x)-D^2_x]h \cdot f=0\,,\\
\label{nyo-71-2}\hspace{-0.8cm}&&[\textmd{i}(D_t-2\alpha_2D_x)+D^2_x]h^* \cdot f=0\,,\\
\hspace{-0.8cm}&& [D_tD_x-2(\sigma_1\rho^2_1+\sigma_2\rho^2_2)]f \cdot f\nonumber\\
 \label{nyo-72}\hspace{-0.8cm}&&\qquad +2\sigma_1\rho^2_1gg^*+2\sigma_2\rho^2_2hh^*=0\,.
\end{eqnarray}
By similar transformations of dependent variables (\ref{nyo-38}),
the above bilinear equations are converted into the 1D coupled YO system.
Thus we have the following theorem about $N$-dark-dark soliton solutions
for the 1D coupled YO system.

\begin{theorem}
The two-component generalization of one-dimensional YO system
\begin{eqnarray}
\label{nyo-73} &&\textmd{i}S^{(1)}_t  - S^{(1)}_{xx}+ L S^{(1)}=0,\\
\label{nyo-74} &&\textmd{i}S^{(2)}_t  - S^{(2)}_{xx}+ L S^{(2)}=0,\\
\label{nyo-75} &&L_t=2(\sigma_1|S^{(1)}|^2+\sigma_2|S^{(2)}|^2)_x,
\end{eqnarray}
has $N$-dark-dark soliton solution:
\begin{eqnarray}
\label{nyo-76}&& S^{(1)}= \rho_1\textmd{e}^{\textmd{i}[\alpha_1x+\alpha^2_1t+\zeta_{10}]}\frac{g}{f},\\
\label{nyo-77}&& S^{(2)}= \rho_2\textmd{e}^{\textmd{i}[\alpha_2x+\alpha^2_2t+\zeta_{20}]}\frac{h}{f},\\
\label{nyo-78}&& L= -2\left(\log f\right)_{xx},
\end{eqnarray}
where $f,g$ and $h$ are Gram determinants given by
\begin{eqnarray*}
&&f=\Bigg|  \delta_{ij} + \frac{1}{p_i+p^*_j} \textmd{e}^{\xi_i+\xi^*_j}  \Bigg|_{N\times N}, \\ &&
  g=\Bigg| \delta_{ij} + \left(-\frac{p_i-\textmd{i}\alpha_1}{p^*_j+\textmd{i}\alpha_1}\right)  \frac{1}{p_i+p^*_j} \textmd{e}^{\xi_i+\xi^*_j} \Bigg|_{N\times N},\ \ \\
&& h=\Bigg| \delta_{ij} + \left(-\frac{p_i-\textmd{i}\alpha_2}{p^*_j+\textmd{i}\alpha_2}\right) \frac{1}{p_i+p^*_j} \textmd{e}^{\xi_i+\xi^*_j}  \Bigg|_{N\times N},\ \
\end{eqnarray*}
where $\xi_j=p_jx - \textmd{i}p^2_jt + \xi_{j0}$, $\alpha_k$, $\rho_k$,
$\zeta_{k0}$
($k{=}1,2$) are real constants, $p_j$, $\xi_{j0}$ are complex constants, and
these parameters  satisfy the constraint conditions:
\begin{eqnarray}
 \frac{\sigma_1\rho^2_1}{|p_j-\textmd{i}\alpha_1|^2}  +  \frac{\sigma_2\rho^2_2}{|p_j-\textmd{i}\alpha_2|^2}  = - \textmd{i} (p_j-p^{*}_j)\,.
\end{eqnarray}
\end{theorem}
\begin{remark}
Compared with the two dimensional case, the
parameters in the $N$-dark-dark
soliton solutions of the 1D coupled YO system need to
satisfy some constraint conditions. In fact, by rewriting the solutions
(\ref{nyo-39})-(\ref{nyo-41}) in
the two dimensional case into the similar forms as (\ref{nyo-58}), one can get
\begin{eqnarray}
\nonumber \hspace{-0.8cm} && \xi_i+\xi^*_i \\
\nonumber\hspace{-0.8cm}&&=  (p_i + p^*_i) \\
\nonumber\hspace{-0.8cm}&&\times \Bigg[x - \bigg( \frac{\sigma_1\rho^2_1}{|p_i-\textmd{i}\alpha_1|^2}
+ \frac{\sigma_2\rho^2_2}{|p_i-\textmd{i}\alpha_2|^2} + \textmd{i}(p_i-p^{*}_i) \bigg)y \\
 \hspace{-0.8cm}&&+\bigg(\frac{\sigma_1\rho^2_1}{|p_i-\textmd{i}\alpha_1|^2}  +\frac{\sigma_2\rho^2_2}{|p_i-\textmd{i}\alpha_2|^2}\bigg)t\Bigg]+ \xi_{i0}+\xi^*_{i0}\,.\label{nyo-79}
\end{eqnarray}
It is easy to find that the constraint conditions in the one-dimensional
case are nothing but the zero condition for the coefficients of $y$ in
the two-dimensional case.
It is interesting that the similar constraint conditions are obtained in
finding the $N$-dark-dark soliton solutions for the coupled NLS equation \cite{ohta2011general}.
\end{remark}

\section{Dynamics of dark-dark solitons}

\subsection{Single dark-dark solitons}
To obtain a single dark-dark soliton solution in Eqs.(\ref{nyo-01})-(\ref{nyo-03}), we take $N=1$ in the formula (\ref{nyo-39})-(\ref{nyo-41}).
The Gram determinants read
\begin{eqnarray}
\label{nyo-80-1}&&f=1{+}\frac{1}{p_1{+}p^*_1} \textmd{e}^{\xi_1+\xi^*_1},\\
\label{nyo-80-2}&&g=1{-}\frac{1}{p_1{+}p^*_1} \frac{p_1- \textmd{i} \alpha_1}{p^*_1 + \textmd{i} \alpha_1} \textmd{e}^{\xi_1+\xi^*_1},\\
\label{nyo-80-3}&&h=1{-}\frac{1}{p_1{+}p^*_1} \frac{p_1- \textmd{i} \alpha_2}{p^*_1 + \textmd{i} \alpha_2} \textmd{e}^{\xi_1+\xi^*_1},
\end{eqnarray}
and then the one-dark-dark soliton solution can be written as
\begin{eqnarray}
\hspace{-0.8cm}&&S^{(1)} = \frac{\rho_1}{2}\textmd{e}^{\textmd{i}[\alpha_1x+\beta_1y-(\beta_1-\alpha^2_1)t+\zeta_{10}]}\nonumber\\
\label{nyo-81}\hspace{-0.8cm}&&\times \bigg[ 1+ K^{(1)}_1 -(1-K^{(1)}_1) \tanh\bigg( \frac{\xi_1+\xi^*_1+\Theta_1}{2}\bigg)\bigg],\\
\hspace{-0.8cm}&&S^{(2)} = \frac{\rho_2}{2}\textmd{e}^{\textmd{i}[\alpha_2x+\beta_2y-(\beta_2-\alpha^2_2)t+\zeta_{20}]}\nonumber\\
\label{nyo-82}\hspace{-0.8cm} &&\times  \bigg[ 1+ K^{(2)}_1
 -(1-K^{(2)}_1) \tanh\bigg(
 \frac{\xi_1+\xi^*_1+\Theta_1}{2}\bigg)\bigg],\\
\label{nyo-83}\hspace{-0.8cm}&& L =-\frac{1}{2}(p_1+p^*_1)^2\rm{sech}^2\bigg( \frac{\xi_1+\xi^*_1+\Theta_1}{2}\bigg),
\end{eqnarray}
with
\begin{eqnarray*}
&& \textmd{e}^{\Theta_1}= \frac{1}{p_1+p^*_1}= \frac{1}{2a_1}, \\
&& K^{(1)}_1=-\frac{p_1- \textmd{i} \alpha_1}{p^*_1 + \textmd{i}
 \alpha_1}=-\frac{a_1 + \textmd{i} (b_1-\alpha_1)}{a_1  -
 \textmd{i}(b_1- \alpha_1)},\\
&&K^{(2)}_1=-\frac{p_1- \textmd{i} \alpha_2}{p^*_1 + \textmd{i} \alpha_2}=-\frac{a_1 + \textmd{i} (b_1-\alpha_2)}{a_1  - \textmd{i}(b_1- \alpha_2)}, \\
&& \xi_1+\xi^*_1=2a_1x \\
&&\quad - \Big(\frac{2\sigma_1a_1\rho^2_1}{a^2_1+(b_1-\alpha_1)^2}+\frac{2\sigma_2a_1\rho^2_2}{a^2_1+(b_1-\alpha_2)^2} -4a_1b_1\Big)y  \\
&&\quad + \Big(\frac{2\sigma_1a_1\rho^2_1}{a^2_1+(b_1-\alpha_1)^2}+\frac{2\sigma_2a_1\rho^2_2}{a^2_1+(b_1-\alpha_2)^2} \Big)t  + 2\xi_{10R},\ \ \
\end{eqnarray*}
where $p_1=a_1+\textmd{i}b_1$, $a_1, b_1, \xi_{10R}, \alpha_i,\beta_i, \rho_i,\zeta_{i0},(i{=}1,2)$ are real constants and $\xi_{10}$ is a complex constant.

\begin{figure*}[!htbp]
\centering
\subfigure[]{\includegraphics[height=1.6in,width=2.2in]{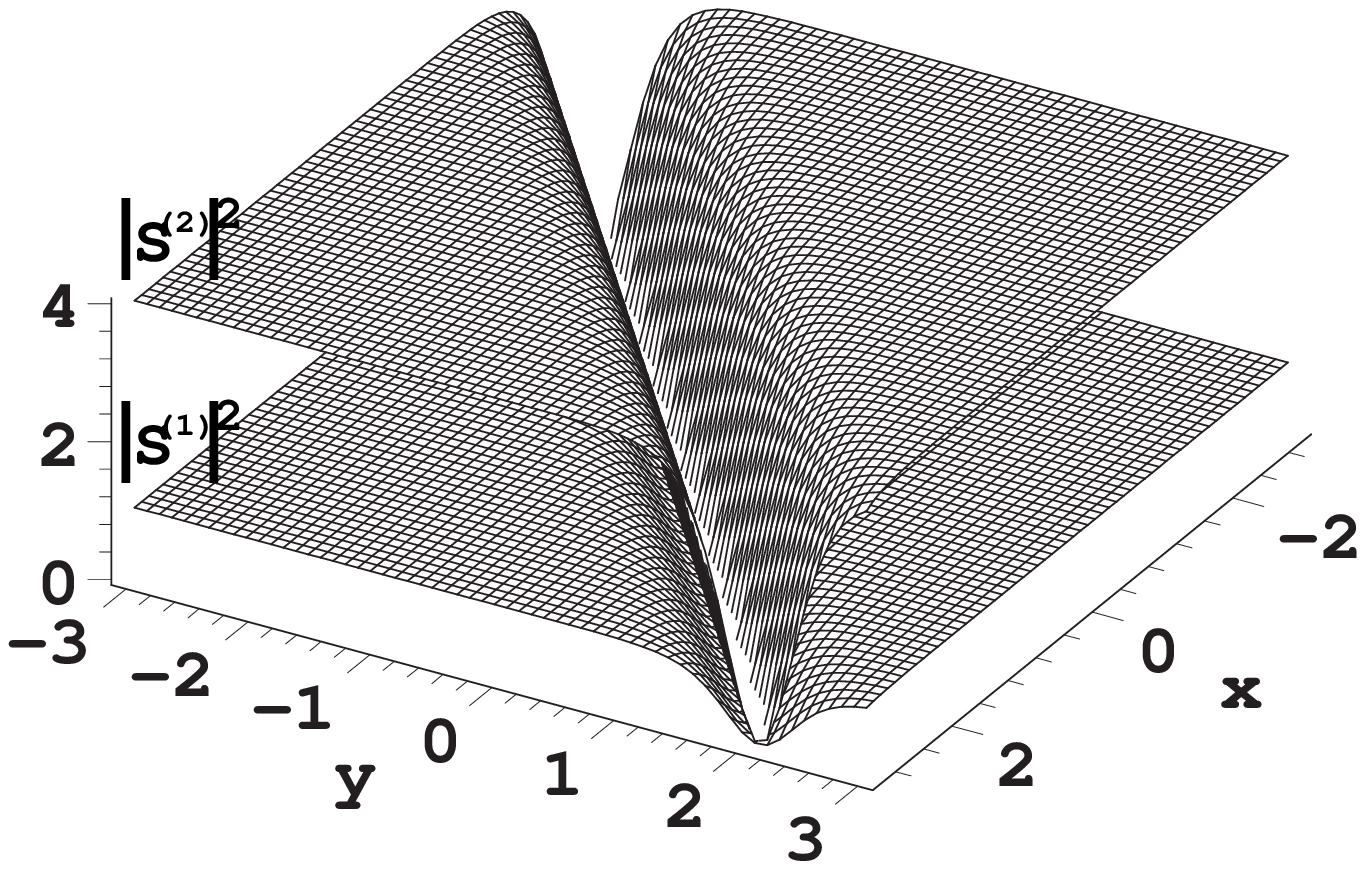}}
\hspace{1cm}
\subfigure[]{\includegraphics[height=1.6in,width=2.2in]{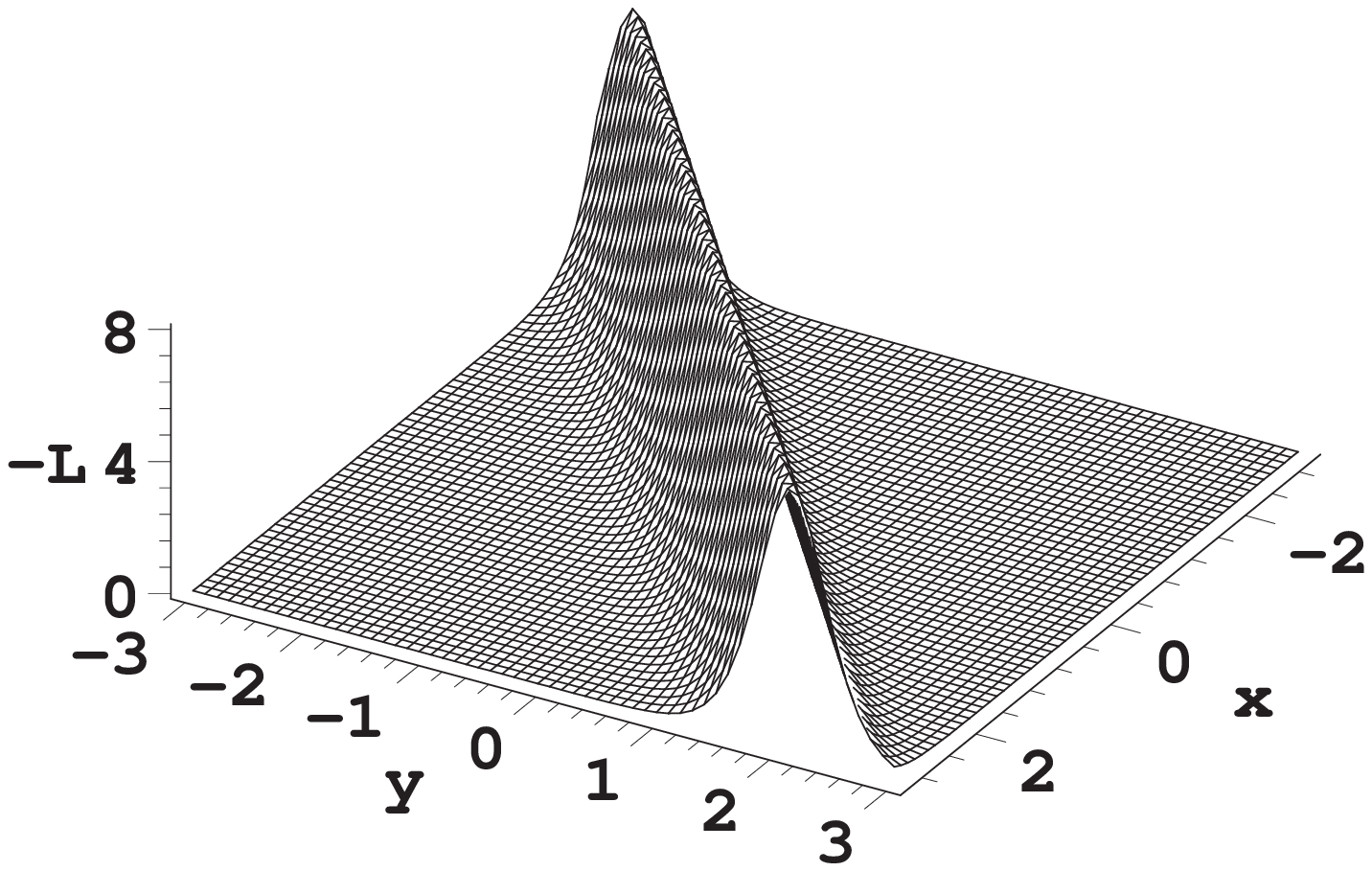}}
\caption{ Single dark-dark solitons (degenerate) at the fixed time $t=0$ with the parameters $\sigma_1=\sigma_2=1,p_1=2, \rho_1=1,\rho_2=2,\alpha_1=\alpha_2=0,\beta_1=1,\beta_2=2$. }
\end{figure*}

\begin{figure*}[!htbp]
\centering
\subfigure[]{\includegraphics[height=1.6in,width=2.2in]{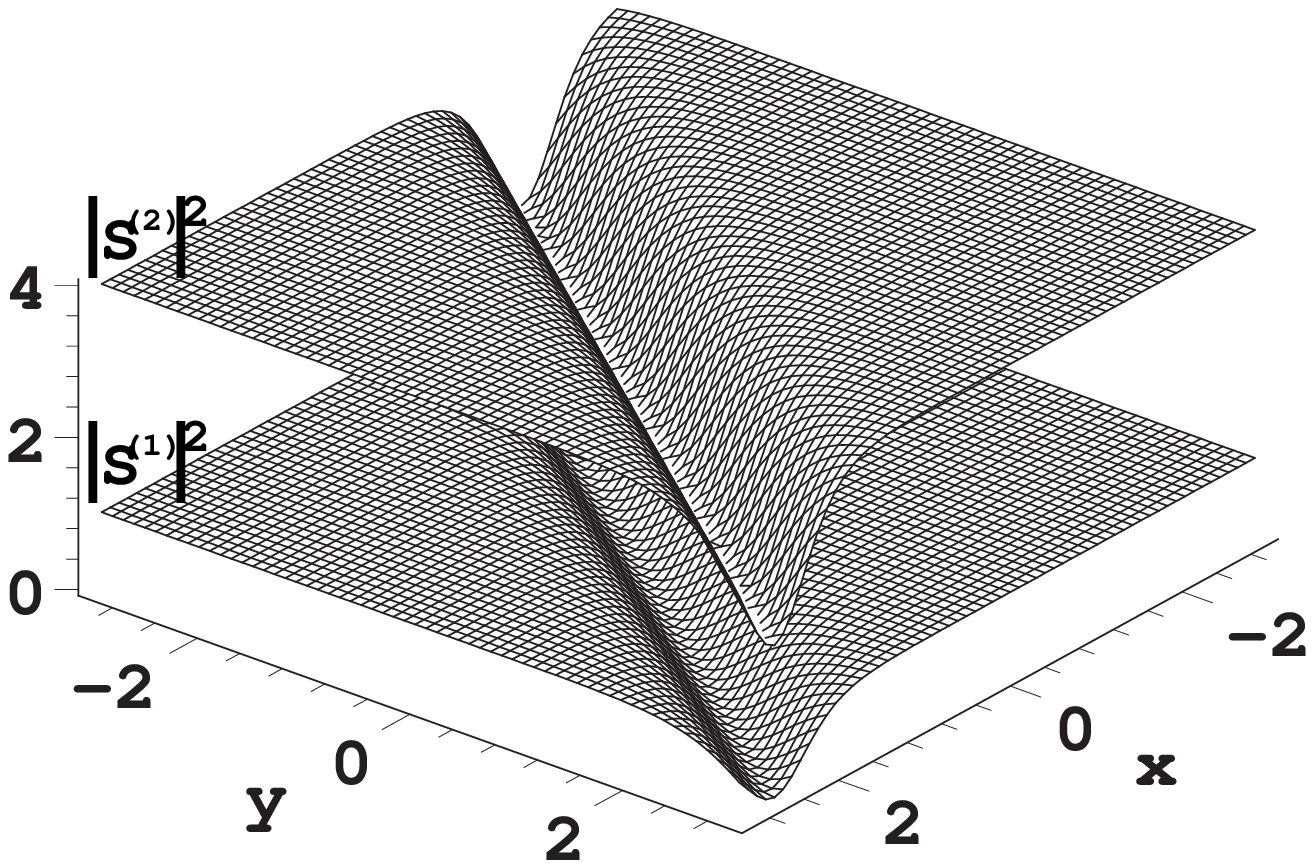}}\hspace{1cm}
\subfigure[]{\includegraphics[height=1.6in,width=2.2in]{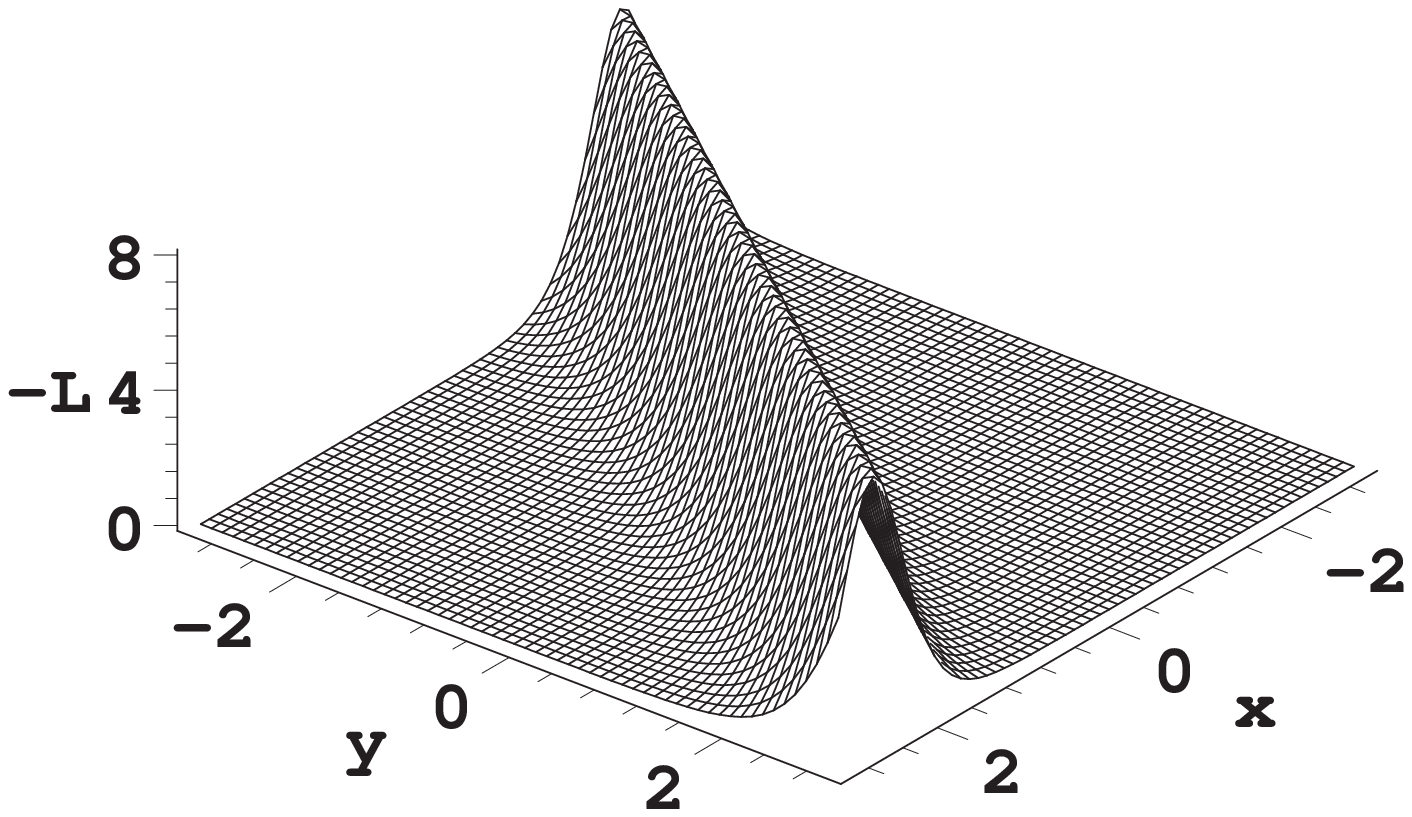}}
\caption{ Single dark-dark solitons (non-degenerate) at the fixed time $t=0$ with the parameters $\sigma_1=\sigma_2=1,p_1=2, \rho_1=1,\rho_2=2,\alpha_1=0,\alpha_2=2,\beta_1=1,\beta_2=2$. }
\end{figure*}

From (\ref{nyo-81})-(\ref{nyo-83}), the intensity functions of the short wave components $|S^{(1)}|, |S^{(2)}|$ and the long-wave component $L$ move at velocity $-\frac{\sigma_1\rho^2_1}{a^2_1+(b_1-\alpha_1)^2}-\frac{\sigma_2\rho^2_2}{a^2_1+(b_1-\alpha_2)^2}$ along the $x$-direction.
As $x,y \rightarrow \pm \infty$, $|S^{(1)}|\rightarrow |\rho_1|, |S^{(2)}|\rightarrow |\rho_2|$, $-L\rightarrow 0$

Denoting $K^{(1)}_1=\exp(2\textmd{i}\phi^{(1)}_1)$ and $K^{(2)}_1=\exp(2\textmd{i}\phi^{(2)}_1)$, the phases of the short wave components $S^{(1)}$ and $S^{(2)}$ acquire shifts in the amount of $2\phi^{(1)}_1$ and $2\phi^{(2)}_1$ but the long wave component $-L$ phase shifts is zero as $x,y$ vary from $-\infty$ to $+\infty$ if $2\phi^{(1)}_1$ and $2\phi^{(2)}_1$ represent the phases of constants $K^{(1)}_1$ and $K^{(1)}_1$ respectively. Without loss of generality, we can assume $2\phi^{(1)}_1, 2\phi^{(2)}_1 \in (-\pi, \pi]$,  (or $\phi^{(1)}_1, \phi^{(2)}_1 \in (-\frac{\pi}{2}, \frac{\pi}{2}]$). Then the intensities of the center of the solitons ($\xi_1+\xi^*_1+\Theta_1=0$) are $|S^{(1)}|_{\small\textrm{center}}=|\rho_1|\cos\phi^{(1)}_1$, $|S^{(2)}|_{\small\textrm{center}}=|\rho_2|\cos\phi^{(2)}_1$ and $-L_{\small\textrm{center}}=2a^2_1$. For the short wave components, the fact that the center intensities are lower than the background intensities implies these solitons are dark-dark solitons.

There are two different cases corresponding to values of $\alpha_1$ and $\alpha_2$:\\
(i) $\alpha_1=\alpha_2$. In this case, $K^{(1)}_1=K^{(2)}_1$, i.e., $\phi^{(1)}_1=\phi^{(2)}_1$, this means
the short wave components $S^{(1)}$ and $S^{(2)}$ are proportional to each other.
In this situation, the dark-dark soliton solution for the coupled YO
system is equivalent to
the dark soliton solution in the single-component YO system, so it is
viewed as degenerate case similar to the coupled NLS equation
\cite{ohta2011general}. We illustrate these degenerate solitons in Fig. 1.\\
(ii) $\alpha_1\neq\alpha_2$. The condition $K^{(1)}_1 \neq K^{(2)}_1$, i.e., $\phi^{(1)}_1 \neq \phi^{(2)}_1$ suggests that
the components $S^{(1)}$ and $S^{(2)}$ have different degrees of
darkness at the center. In this non-degenerate single dark-dark solitons
of the coupled YO system (\ref{nyo-01})-(\ref{nyo-03}), the components
$S^{(1)}$ and $S^{(2)}$ are not proportional to each other. As is shown
in Fig. 2, the intensity of the component $S^{(1)}$  is black, but the intensity of the component $S^{(2)}$  is gray at their centers.


\subsection{Two-dark-dark solitons}
The two-dark-dark soliton solution can be obtained by taking $N=2$ in the formula (\ref{nyo-39})-(\ref{nyo-41}).
In this case, we have
\begin{eqnarray}
\label{nyo-85} && S^{(1)}= \rho_1\textmd{e}^{\textmd{i}[\alpha_1x+\beta_1y-(\beta_1-\alpha^2_1)t+\zeta_{10}]}\frac{g_2}{f_2},\\
\label{nyo-86} && S^{(2)}= \rho_2\textmd{e}^{\textmd{i}[\alpha_2x+\beta_2y-(\beta_2-\alpha^2_2)t+\zeta_{20}]}\frac{h_2}{f_2},\\
\label{nyo-87} && L= -2\left(\log f_2\right)_{xx},
\end{eqnarray}
with
\begin{eqnarray}
\hspace{-0.8cm}&& f_2=1+
 \textmd{e}^{\xi_1+\xi^*_1+\Theta_1} +
 \textmd{e}^{\xi_2+\xi^*_2+\Theta_2} \\
\hspace{-0.8cm}&&\qquad \qquad  + \Omega_{12}
 \textmd{e}^{\xi_1+\xi^*_1+\xi_2+\xi^*_2+\Theta_1+\Theta_2},\label{nyo-88} \\
\hspace{-0.8cm}\label{nyo-89} && g_2=1+
 K^{(1)}_1\textmd{e}^{\xi_1+\xi^*_1+\Theta_1} +  K^{(1)}_2
 \textmd{e}^{\xi_2+\xi^*_2+\Theta_2} \nonumber\\
\hspace{-0.8cm}\label{nyo-89}&&\qquad \qquad
+ \Omega_{12}  K^{(1)}_1  K^{(1)}_2 \textmd{e}^{\xi_1+\xi^*_1+\xi_2+\xi^*_2+\Theta_1+\Theta_2},\\
\hspace{-0.8cm} && h_2=1+
 K^{(2)}_1\textmd{e}^{\xi_1+\xi^*_1+\Theta_1} +  K^{(2)}_2
 \textmd{e}^{\xi_2+\xi^*_2+\Theta_2} \nonumber\\
\hspace{-0.8cm}\label{nyo-90} &&
\qquad \qquad  + \Omega_{12}  K^{(2)}_1  K^{(2)}_2 \textmd{e}^{\xi_1+\xi^*_1+\xi_2+\xi^*_2+\Theta_1+\Theta_2},
\end{eqnarray}
and
\begin{eqnarray*}
\hspace{-0.8cm}&& \textmd{e}^{\Theta_j}= \frac{1}{p_j+p^*_j}=\frac{1}{2a_j},\\
\hspace{-0.8cm}&& \Omega_{12}=\bigg|\frac{p_1-p_2}{p_1+p^*_2}\bigg|^2=
\bigg|\frac{a_1-a_2+\textmd{i}(b_1-b_2)}{a_1+a_2+\textmd{i}(b_1-b_2)}\bigg|^2
=\frac{(a_1-a_2)^2+(b_1-b_2)^2}{(a_1+a_2)^2+(b_1-b_2)^2}
,\\
\hspace{-0.8cm}&& K^{(1)}_j=-\frac{p_j- \textmd{i} \alpha_1}{p^*_j + \textmd{i} \alpha_1}=-\frac{a_j + \textmd{i} (b_j-\alpha_1)}{a_j  - \textmd{i}(b_j- \alpha_1)},\\
\hspace{-0.8cm}&& K^{(2)}_j=-\frac{p_j- \textmd{i} \alpha_2}{p^*_j + \textmd{i} \alpha_2}=-\frac{a_j + \textmd{i} (b_j-\alpha_2)}{a_j  - \textmd{i}(b_j- \alpha_2)},\\
\hspace{-0.8cm}  && \xi_j+\xi^*_j= k_{x,j} x+k_{y,j} y+\omega_j t+ 2\xi_{j0R},\\
\hspace{-0.8cm}  &&\quad = 2a_jx \\
\hspace{-0.8cm}&&\qquad  - \Big(\frac{2\sigma_1a_j\rho^2_1}{a^2_j+(b_j-\alpha_1)^2}+\frac{2\sigma_2a_j\rho^2_2}{a^2_j+(b_j-\alpha_2)^2} -4a_jb_j\Big)y  \\
\hspace{-0.8cm}&&\qquad
+ \Big(\frac{2\sigma_1a_j\rho^2_1}{a^2_j+(b_j-\alpha_1)^2}+\frac{2\sigma_2a_j\rho^2_2}{a^2_j+(b_j-\alpha_2)^2} \Big)t  + 2\xi_{j0R},\ \ \
\end{eqnarray*}
where $p_j=a_j+\textmd{i}b_j$,  $a_j,b_j,\alpha_j,\beta_j, \rho_j,\zeta_{j0},(j{=}1,2)$ are real constants, and $\xi_{10},\xi_{20}$ are complex constants.

\begin{remark}
In the case of $a_2=-a_1$ and $b_2=b_1$ (i.e., $p_2=-p_1^*$),
the denominator of $\Omega_{12}$ becomes zero.
On this critical wave number, the soliton interaction shows Y-shape.
This Y-shape type soliton solution is called the resonant soliton
solution which was found in the KP equation.
As the two-soliton solution of the KP equation,
the above two-soliton solution are classified into two
different types of soliton interactions~\cite{CK:08,CK:09,K:10,CLM:10}:
\begin{enumerate}
 \item If $a_1a_2<0$, $1<\Omega_{12}$. This case is called the O-type
 soliton interaction. In this case, two asymptotic soliton amplitudes
$2a_1^2$ and $2a_2^2$ (in the variable $-L$) can be equivalent when $a_2=-a_1$.
The interaction peak (the maximum of $\-L$) is
always greater than the sum of the asymptotic soliton amplitudes.

\item If $a_1a_2>0$, $0<\Omega_{12}<1$. This case is called the P-type
 soliton interaction. In this case, two asymptotic soliton amplitudes
$2a_1^2$ and $2a_2^2$ (in the variable $-L$) cannot be
 equivalent.
The interaction peak (the maximum of $-L$) is
always less than the sum of the asymptotic soliton amplitudes.
\end{enumerate}

Note that types of soliton interactions do not depend on the parameters
$b_1$ and $b_2$ (i.e., the imaginary parts of $p_1$ and $p_2$).
The resonant Y-shape soliton solution is obtained by taking
the limit $b_2\to b_1$ in the equal-amplitude O-type two-soliton
 ($a_2=-a_1$).
The interaction coefficient $\Omega_{12}$ for the two-soliton solution
of the 2D coupled YO system is always non-negative
although it can be negative for the KP two-soliton solution.
\end{remark}

\begin{figure*}[!htbp]
\centering
\subfigure[]{\includegraphics[height=1.4in,width=1.9in]{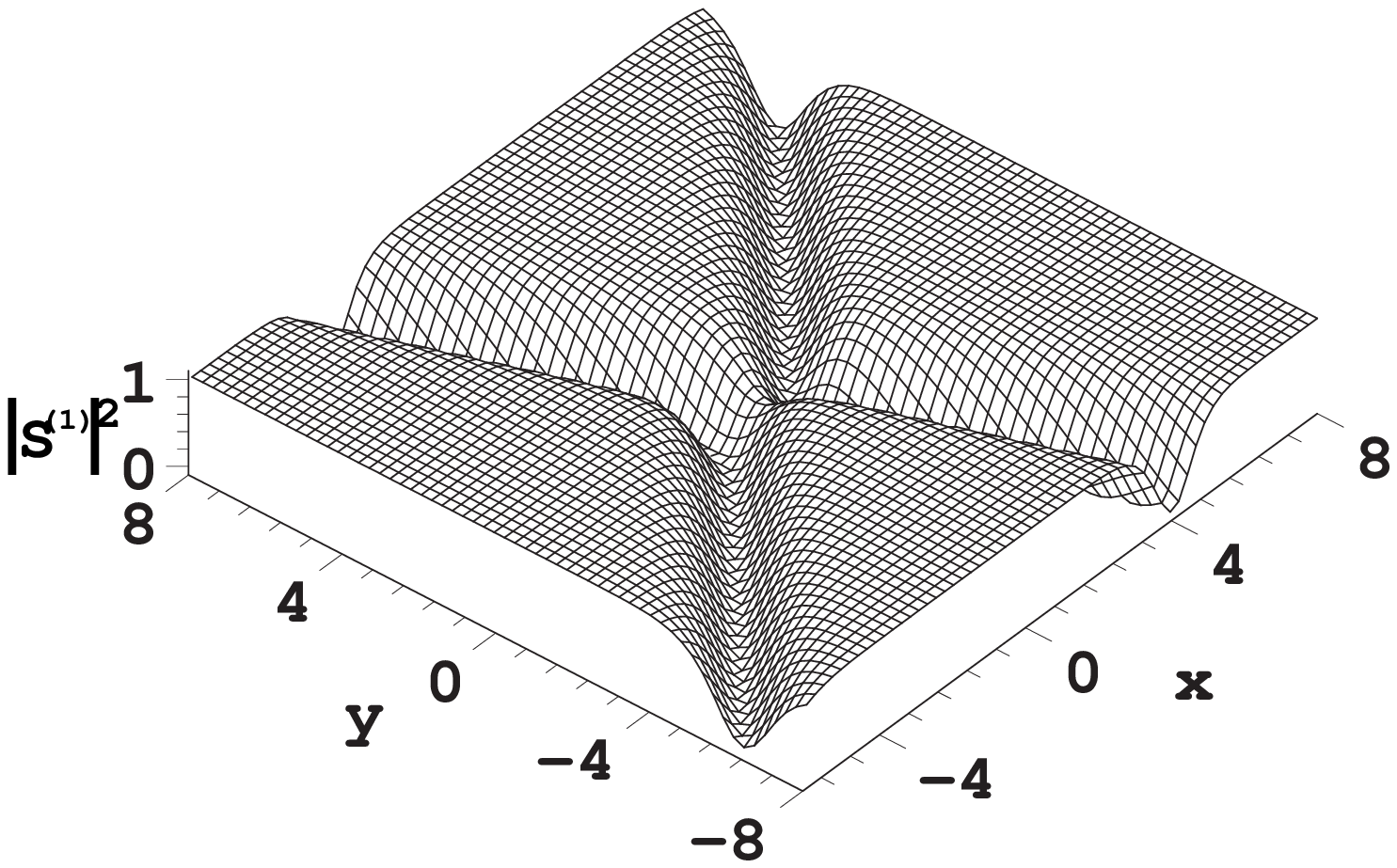}}\hspace{0.5cm}
\subfigure[]{\includegraphics[height=1.4in,width=1.9in]{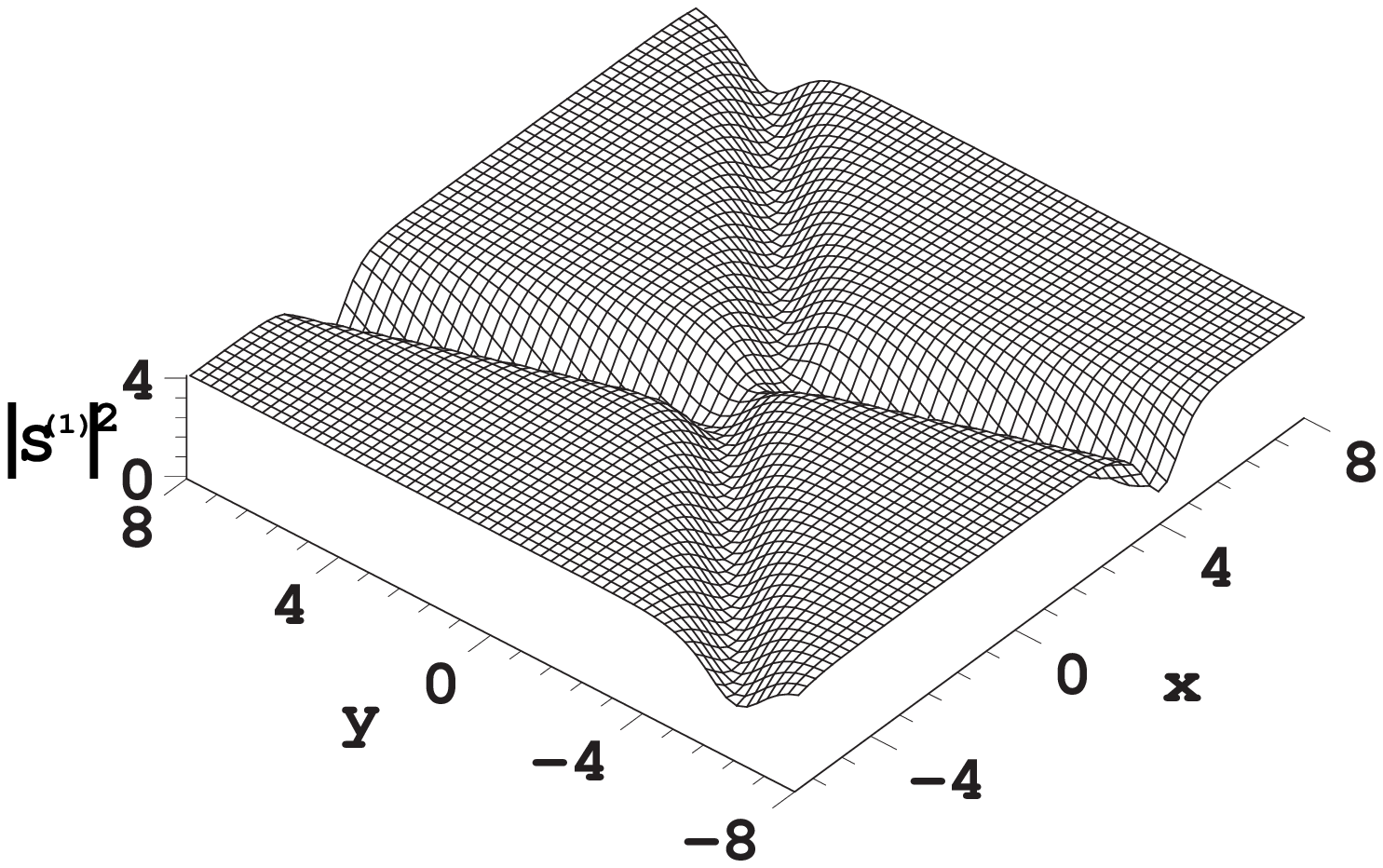}}\hspace{0.5cm}
\subfigure[]{\includegraphics[height=1.4in,width=1.9in]{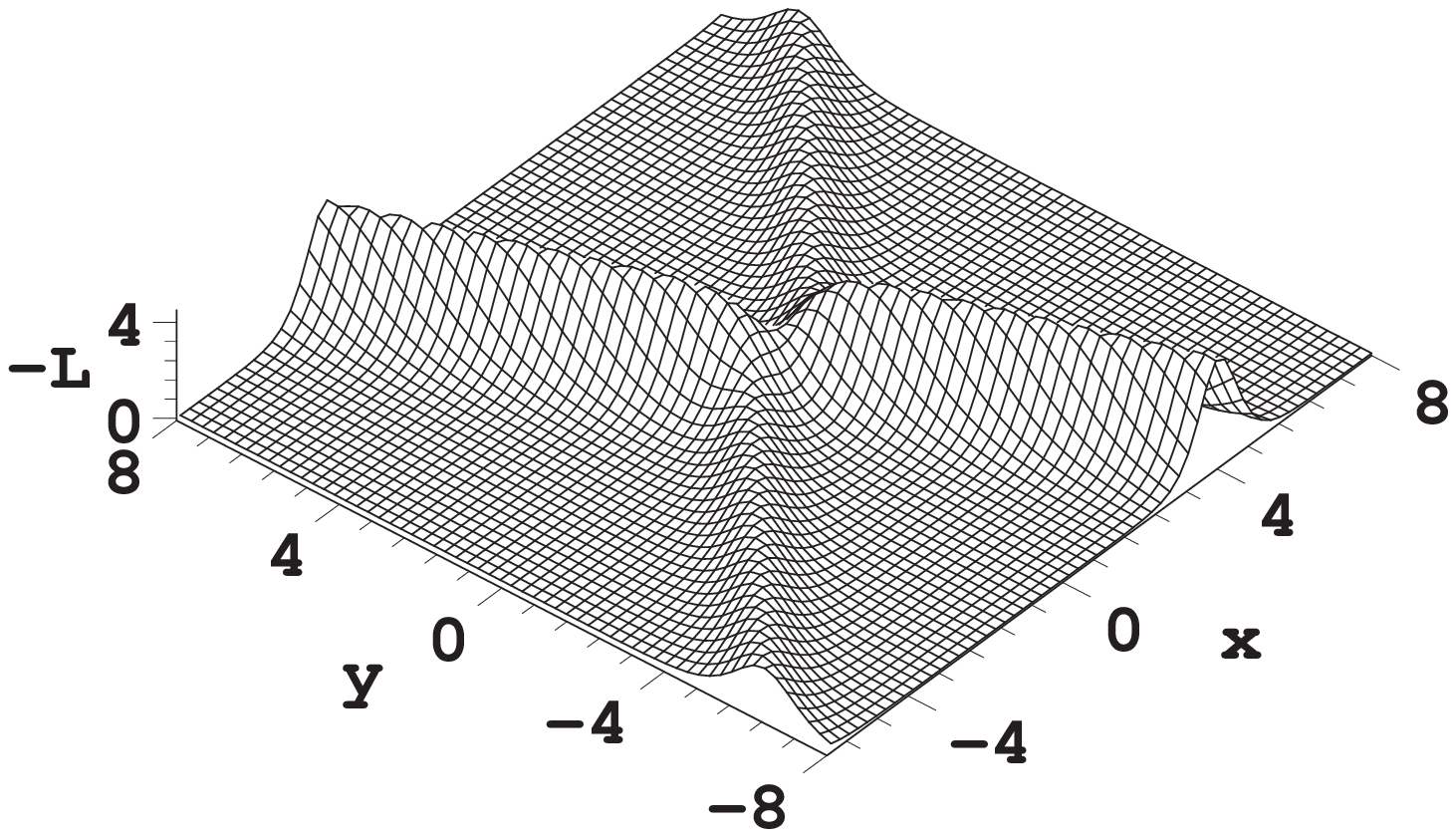}}
\caption{ Two dark-dark solitons at the fixed time $t=0$ the parameters $\sigma_1=\sigma_2=1,p_1=1+\textmd{i}, p_2=1.5+\textmd{i}, \rho_1=1,\rho_2=2,\alpha_1=1,\alpha_2=2,\beta_1=1,\beta_2=2$. }
\end{figure*}

The collision of two dark-dark solitons is displayed in Fig. 3. It is
easy to observe that
the two solitons pass through each
other without any change of shape, darkness and velocity in both
components after collision.
Hence there is no energy transfer between the two solitons or between the $S^{(1)}$
and $S^{(2)}$ components after collision. This complete transmission of
energy of dark-dark
soliton in both components occurs not only for $\sigma_1=\sigma_2=1$ as
in Fig. 3, but also for all other $\sigma_1$ and $\sigma_2$ values. For
the coupled YO system, this kind of phenomenon is distinctly different
from collisions of bright-bright solitons. As reported in the paper by
Kanna, Vijayajayanthi, Sakkaravarthi and
Lakshmanan\cite{kanna2009higher},
the bright-bright solitons in the short wave components $S^{(1)}$ and
$S^{(2)}$
undergo shape changing (energy redistribution) collisions while the long wave component only have an elastic collision.

\section{Dark-dark soliton bound states}

In this section, we investigate the soliton bound states.
To obtain two dark-dark soliton bound states of the coupled YO system, the parameters need to satisfy  $\frac{\omega_1}{k_{x,1}}=\frac{\omega_2}{k_{x,2}}$ and $\frac{\omega_1}{k_{y,1}}=\frac{\omega_2}{k_{y,2}}$,
which results in two solitons with the same velocity in both short and long wave components.


\subsection{The stationary dark-dark soliton bound states}
The stationary dark-dark soliton bound states means that the common
 velocity equals zero. The stationary solitons for the 2D coupled YO
 system are possible when $\sigma_1$ and $\sigma_2$ take opposite
 signs. Requiring the coefficients of $t$ in the solution
 (\ref{nyo-39})-(\ref{nyo-41}) of the 2D coupled YO system
 ($\sigma_1=1,\sigma_2=-1$) to be zero, i.e.,
 $\frac{\rho^2_1}{|p_i-\textmd{i}\alpha_1|^2}  -
 \frac{\rho^2_2}{|p_i-\textmd{i}\alpha_2|^2}=0$. The degenerate case
 ($\alpha_1=\alpha_2$) leads to $\rho^2_1=\rho^2_2$, the dark soliton
 solutions for two short wave components are equivalent. The
 non-degenerate situation $\alpha_1 \neq \alpha_2$ can be further
 divided into two subcases ($p_j=a_j+\textmd{i}b_j$):
\begin{itemize}
 \item Case (a) $\rho^2_1 = \rho^2_2$: We have
       $b_j=\frac{\alpha_1+\alpha_2}{2}$ and $K^{(1)}_j=K^{(2)*}_j$. This case is trivial.
\item Case (b) $\rho^2_1\neq \rho^2_2$: $a_j=\sqrt{\frac{\rho^2_2(b_j-\alpha_1)^2-\rho^2_1(b_j-\alpha_2)^2}{\rho^2_1-\rho^2_2}}$.
\end{itemize}
In case (b), the soliton solution is independent of the time $t$ and
$|S^{(1)}|^2-|S^{(2)}|^2=\rho^2_1-\rho^2_2$.
Thus the original YO system reduces to two component linear Schr\"{o}dinger equations with potential $L(x,y)$ if $y$ is viewed as the time variable.
That is to say, the linear Schr\"{o}dinger equation
\begin{eqnarray}\label{nyo-84}
\textmd{i}S_y - S_{xx}+ L S=0,
\end{eqnarray}
possess two dark soliton solutions expressed by the form
(\ref{nyo-39})-(\ref{nyo-41}) with the constraints
$\frac{\rho^2_1}{|p_i-\textmd{i}\alpha_1|^2}  -
\frac{\rho^2_2}{|p_i-\textmd{i}\alpha_2|^2}=0$ and $\beta_k=\alpha^2_k$
for $k=1,2$.

Two examples of bound states are illustrated in Fig. 4 and Fig. 5,
respectively. Fig. 4 shows a case of $\frac{k_{y,1}}{k_{x,1}} \neq
\frac{k_{y,2}}{k_{x,2}}$, which corresponds to an oblique bound state. Whereas, Fig. 5 displays a trivial case of
 $\frac{k_{y,1}}{k_{x,1}}=\frac{k_{y,2}}{k_{x,2}}$, which corresponds to a quasi-one-dimensional one. This
kind of bound states can be viewed as two dark-dark soliton bound states of the linear Schr\"{o}dinger equation (\ref{nyo-84}) with potential $L$.

\begin{figure*}[!htbp]
\centering
\subfigure[]{\includegraphics[height=1.4in,width=1.9in]{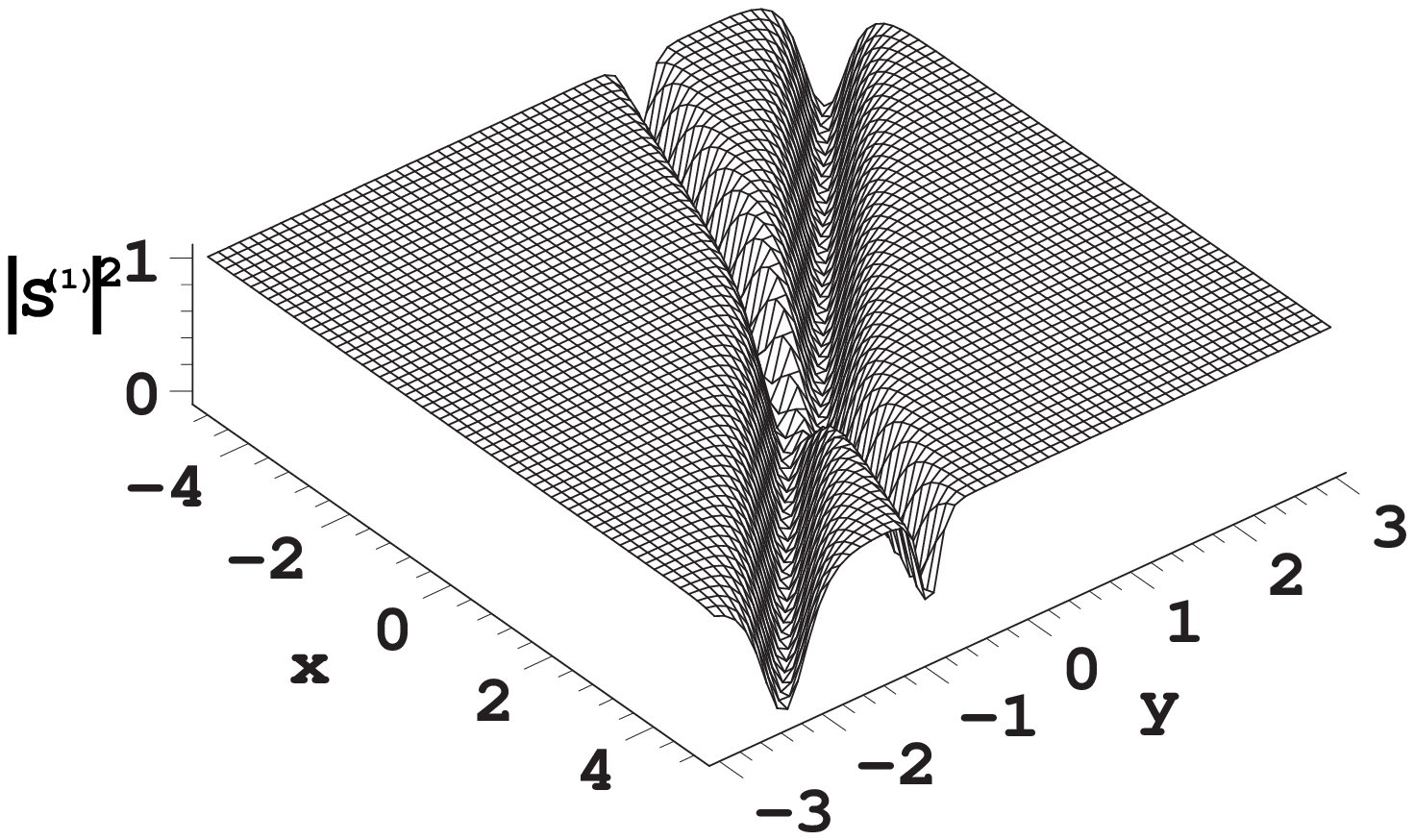}}\hspace{0.5cm}
\subfigure[]{\includegraphics[height=1.4in,width=1.9in]{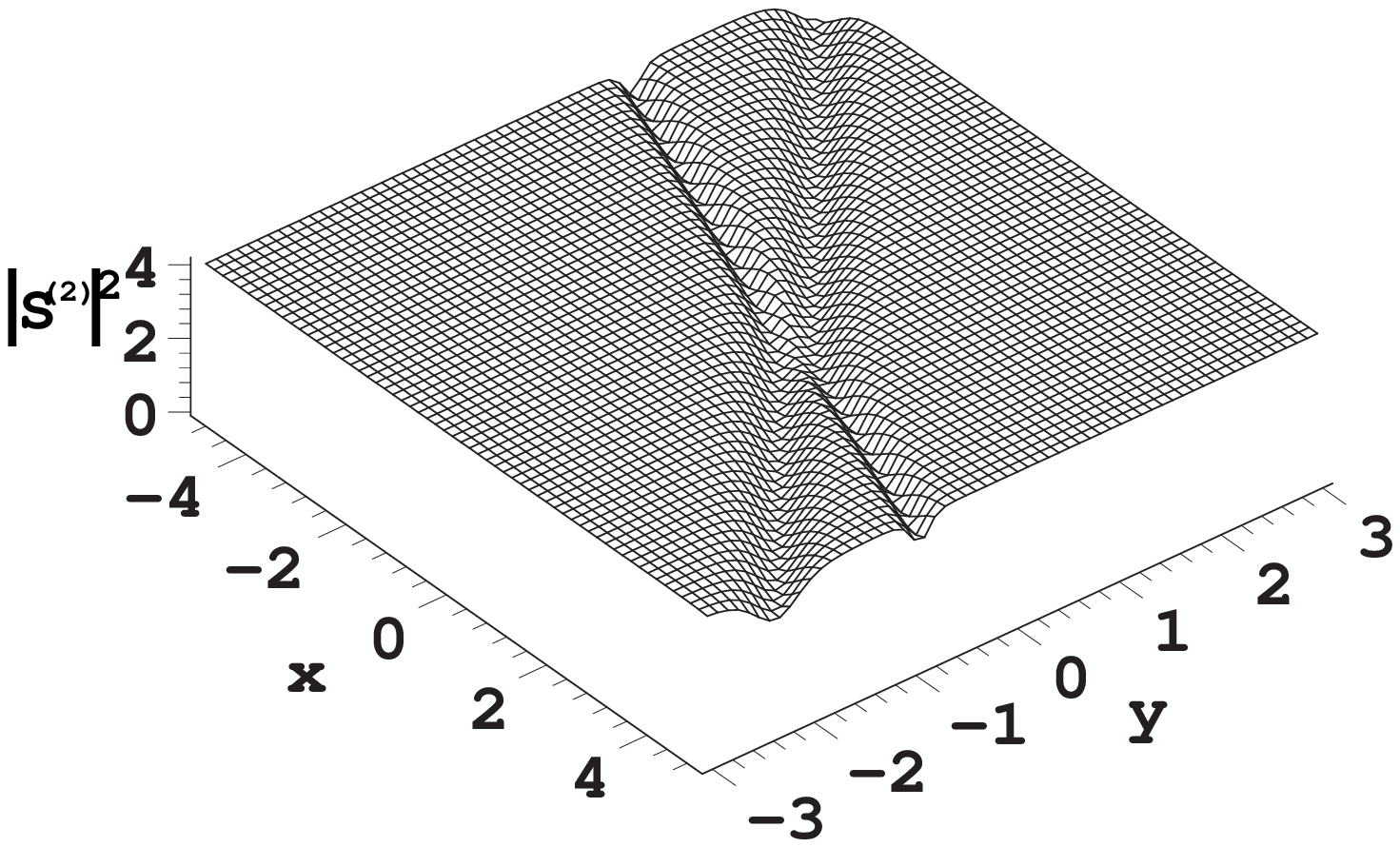}}\hspace{0.5cm}
\subfigure[]{\includegraphics[height=1.4in,width=1.9in]{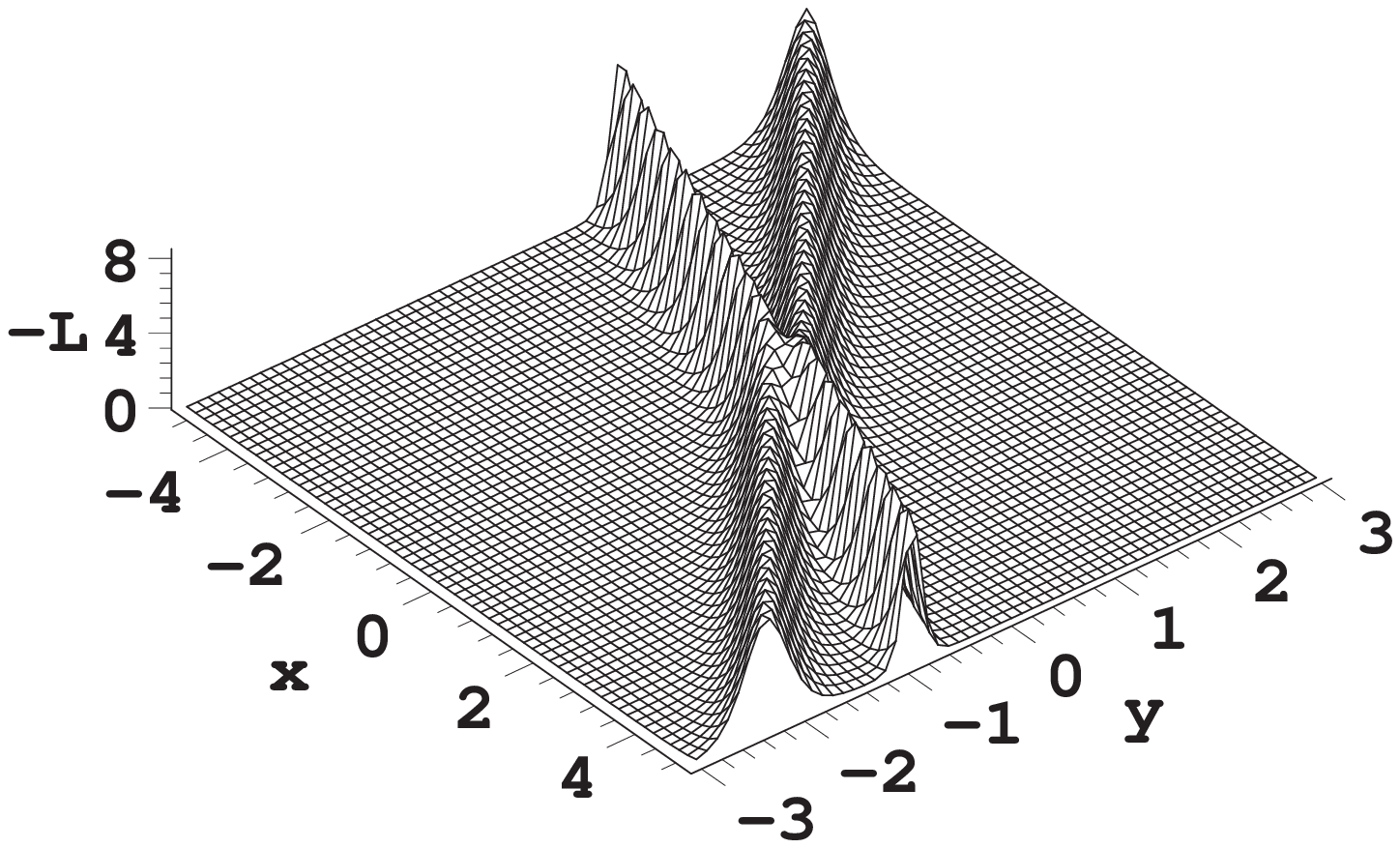}}
\caption{ The stationary dark-dark soliton bound states under the
 condition
$\frac{k_{y,1}}{k_{x,1}}\neq \frac{k_{y,2}}{k_{x,2}}$
with the parameters $\sigma_1=1,\sigma_2=-1,p_1=\sqrt{3}+\textmd{i}, p_2=\frac{9}{4}+\frac{3\sqrt{7}}{4}\textmd{i}, \rho_1=1,\rho_2=2,\alpha_1=1,\alpha_2=-2,\beta_1=1,\beta_2=2$. }
\end{figure*}

\begin{figure*}[!htbp]
\centering
\subfigure[]{\includegraphics[height=1.4in,width=1.9in]{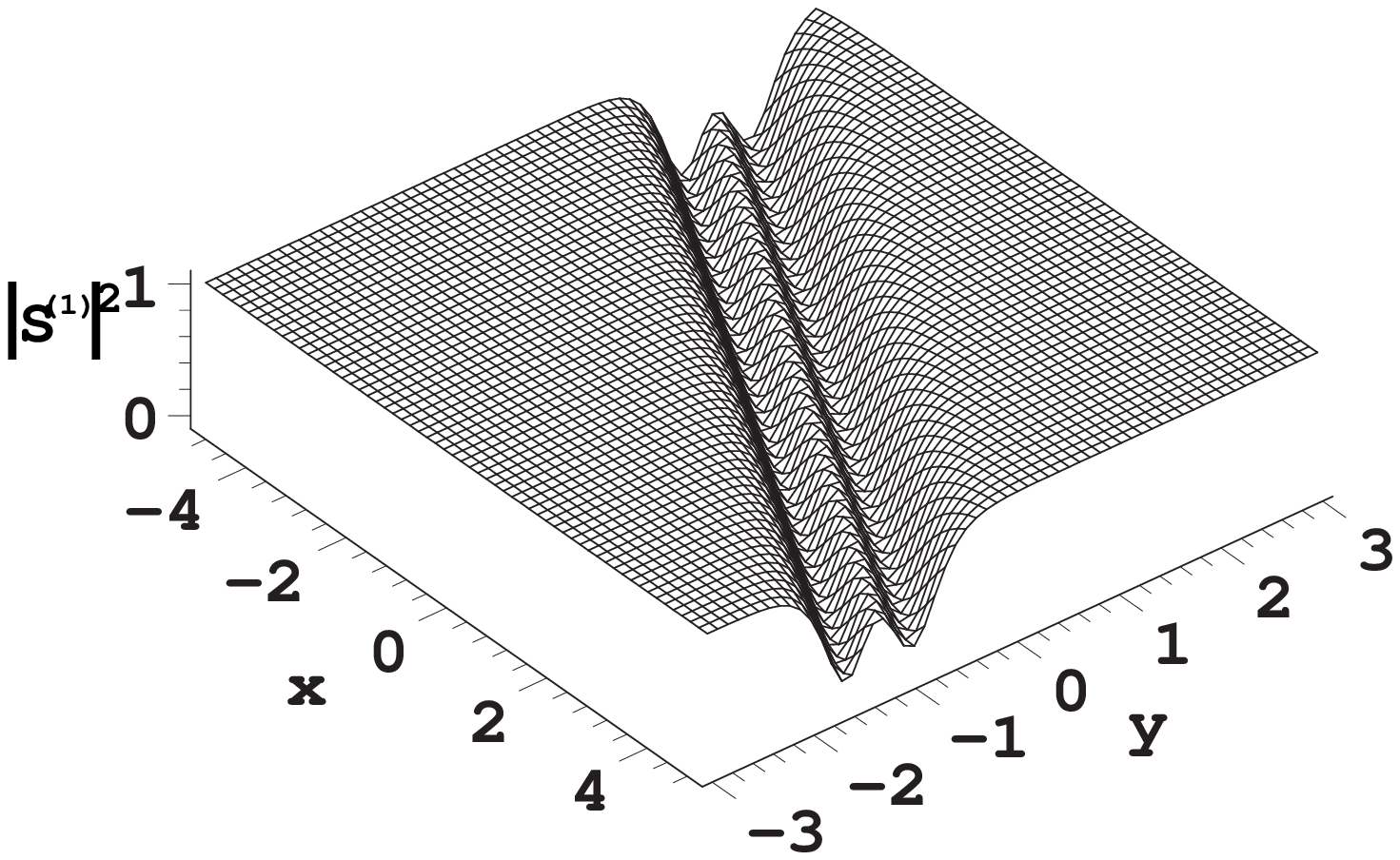}}\hspace{0.5cm}
\subfigure[]{\includegraphics[height=1.4in,width=1.9in]{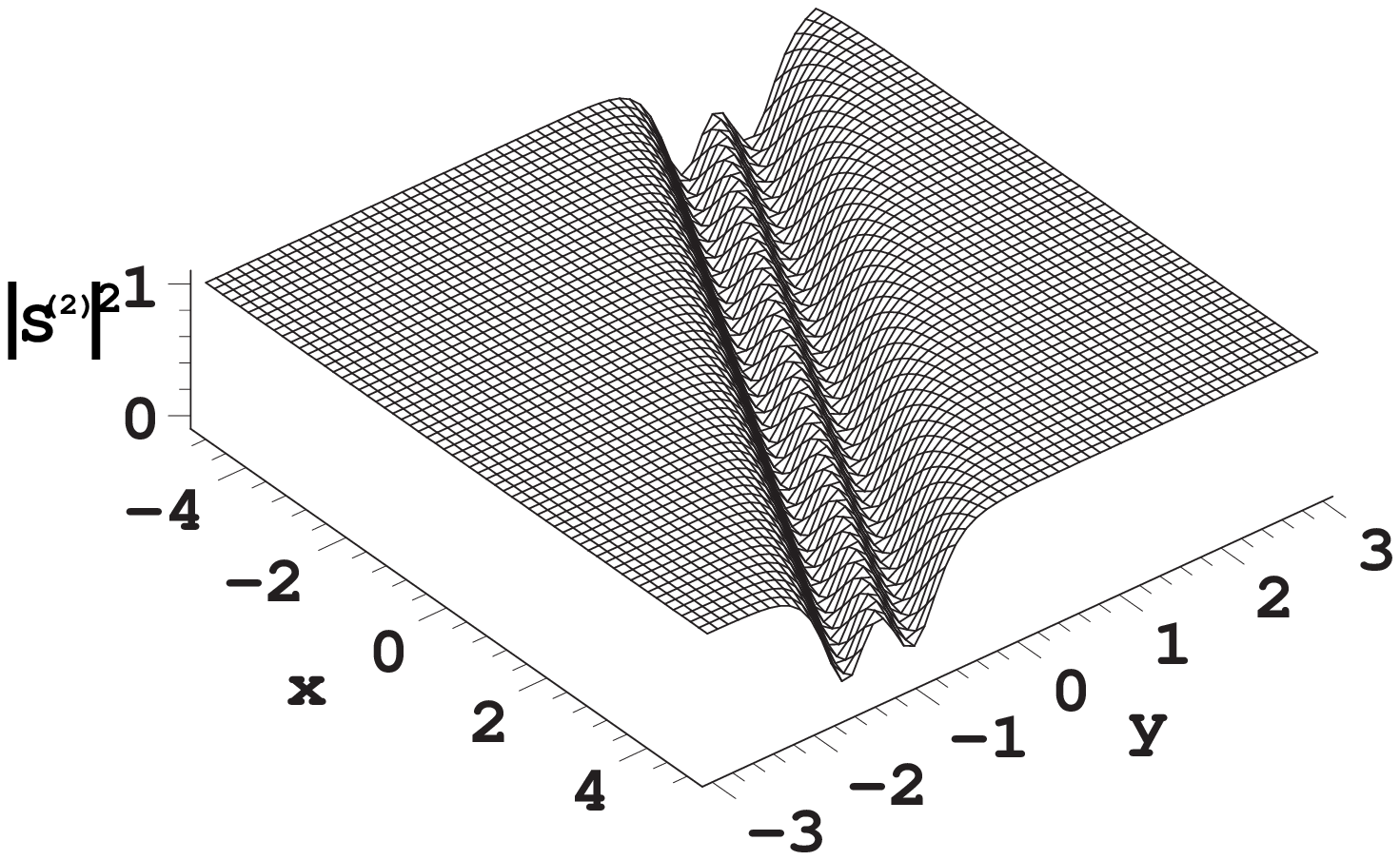}}\hspace{0.5cm}
\subfigure[]{\includegraphics[height=1.4in,width=1.9in]{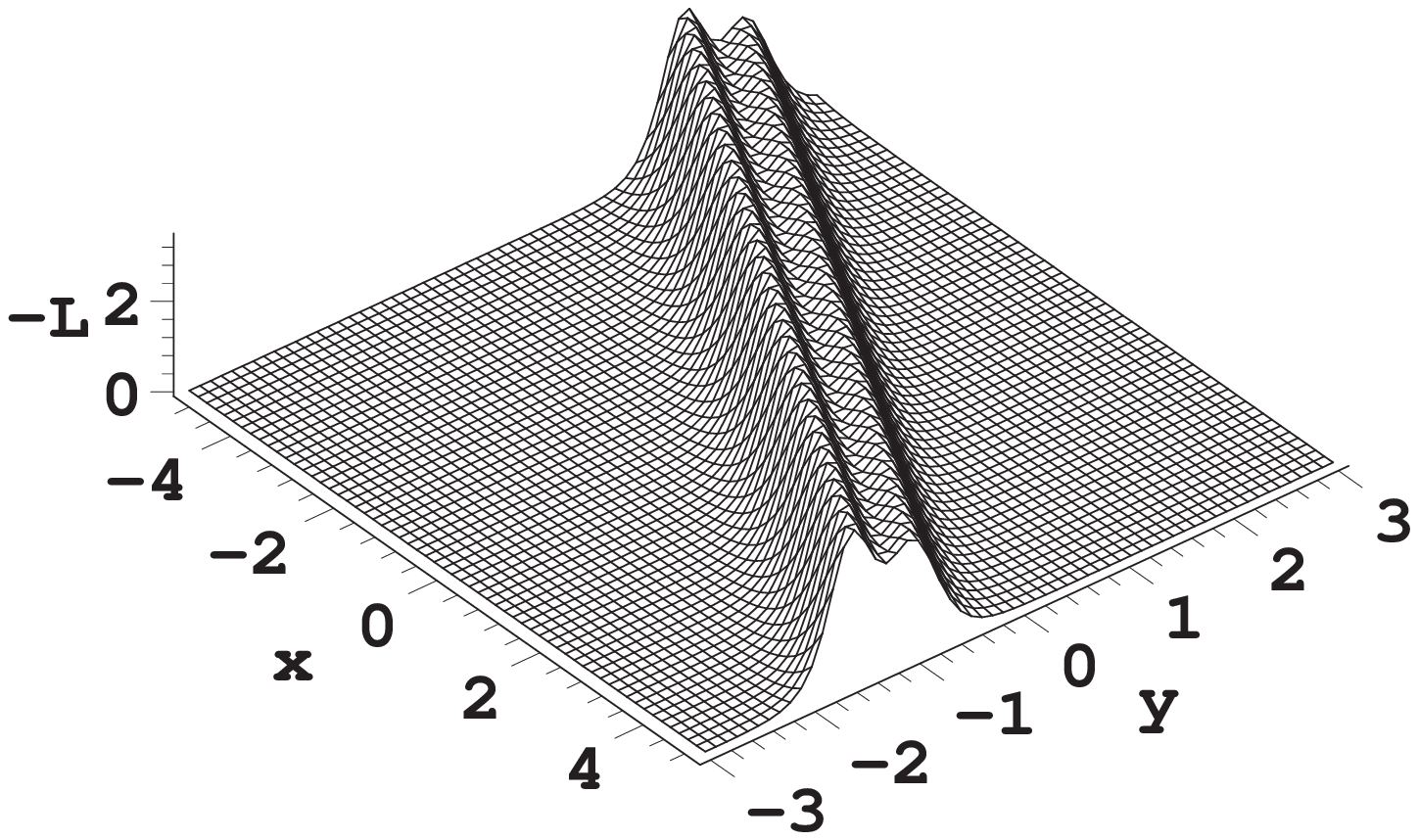}}
\caption{ The stationary dark-dark soliton bound states under the
 condition
$\frac{k_{y,1}}{k_{x,1}}= \frac{k_{y,2}}{k_{x,2}}$
with the parameters $\sigma_1=1,\sigma_2=-1,p_1=1+ \frac{3}{2}\textmd{i}, p_2=\frac{3}{2}+\frac{3}{2}\textmd{i}, \rho_1=1,\rho_2=1,\alpha_1=1,\alpha_2=2,\beta_1=1,\beta_2=2$.  }
\end{figure*}

\subsection{The moving bound dark-dark soliton states}

The moving bound dark-dark soliton states require the common velocity being nonzero, i.e.,
$\omega_1\neq 0$ and $\omega_2\neq0$.
Then the parameters need to satisfy the following condition:
\begin{eqnarray}
\hspace{-0.8cm}&& b_1=b_2\,,\nonumber \\
\hspace{-0.8cm}&& a_2=\sqrt{-\frac{\sigma_1\rho^2_1\alpha^{'2}_2(a^2_1+\alpha^{'2}_2)+\sigma_2\rho^2_2\alpha^{'2}_1(a^2_1+\alpha^{'2}_1)}
{\sigma_1\rho^2_1(a^2_1+\alpha^{'2}_2)+\sigma_2\rho^2_2(a^2_1+\alpha^{'2}_1)}}\,,\nonumber\\
&&\label{nyo-91}
\end{eqnarray}
where $\alpha'_1=b_1-\alpha_1$ and $\alpha'_2=b_2-\alpha_2$. From the above expression, $\sigma_1$ and $\sigma_2$ must take different signs. To show these moving bound dark-dark soliton states, we choose the parameters as
\begin{eqnarray}
\hspace{-0.8cm}&&
\sigma_1{=}-\sigma_2{=}\alpha_1{=}\beta_1{=}\rho_1{=}b_1{=}b_2{=}a_1{=}1,\nonumber\\
\hspace{-0.8cm}&&\alpha_2=\frac{1}{4},\ \  \beta_2{=}2,\ \ a_2{=}\frac{3}{2},\ \ \rho_2{=}\frac{5\sqrt{5}}{8},\label{nyo-92}
\end{eqnarray}
and the corresponding profiles are displayed in Figs. 6--8 at different times.

\begin{figure*}[!htbp]
\centering
\subfigure[]{\includegraphics[height=1.4in,width=1.9in]{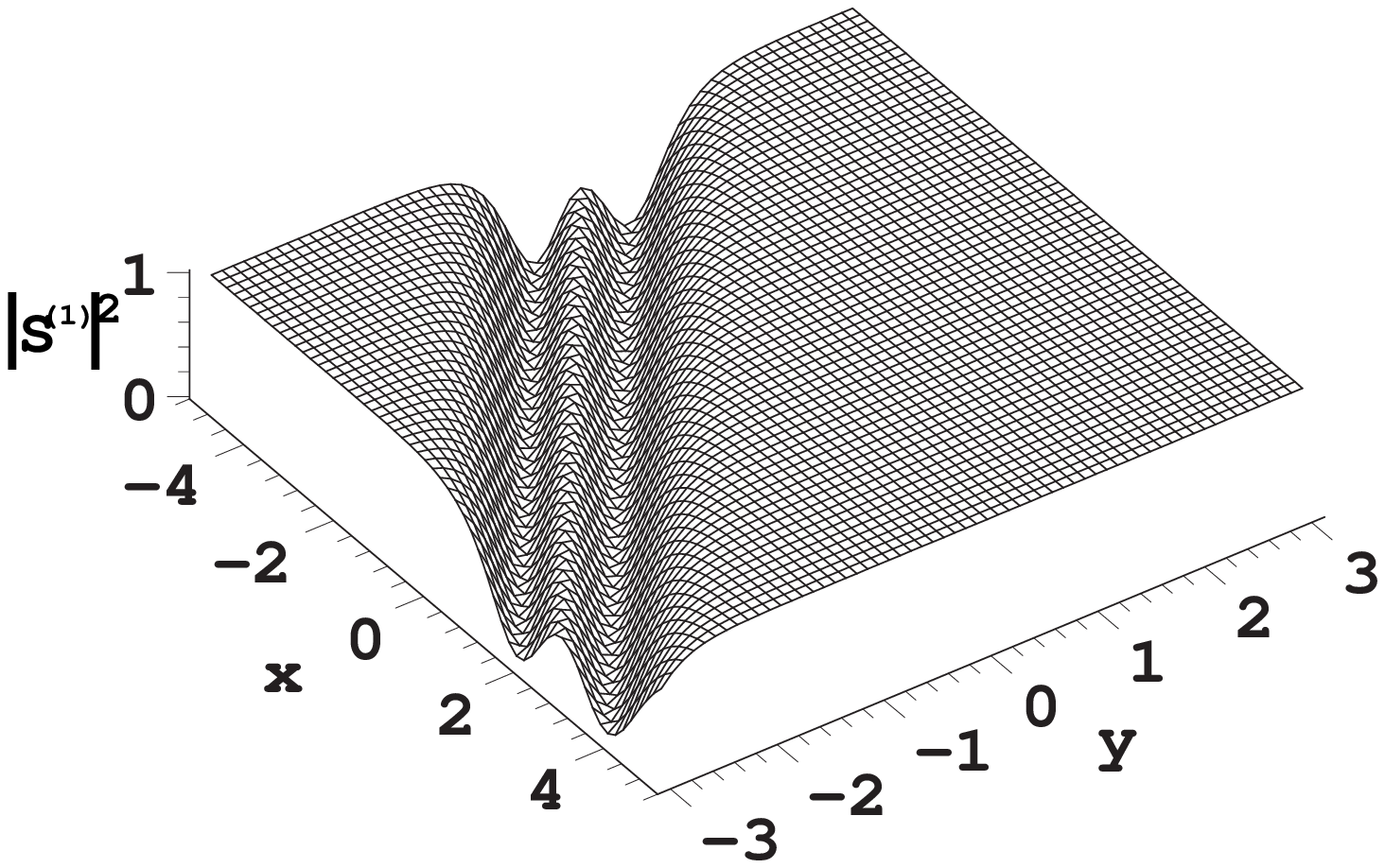}}\hspace{0.5cm}
\subfigure[]{\includegraphics[height=1.4in,width=1.9in]{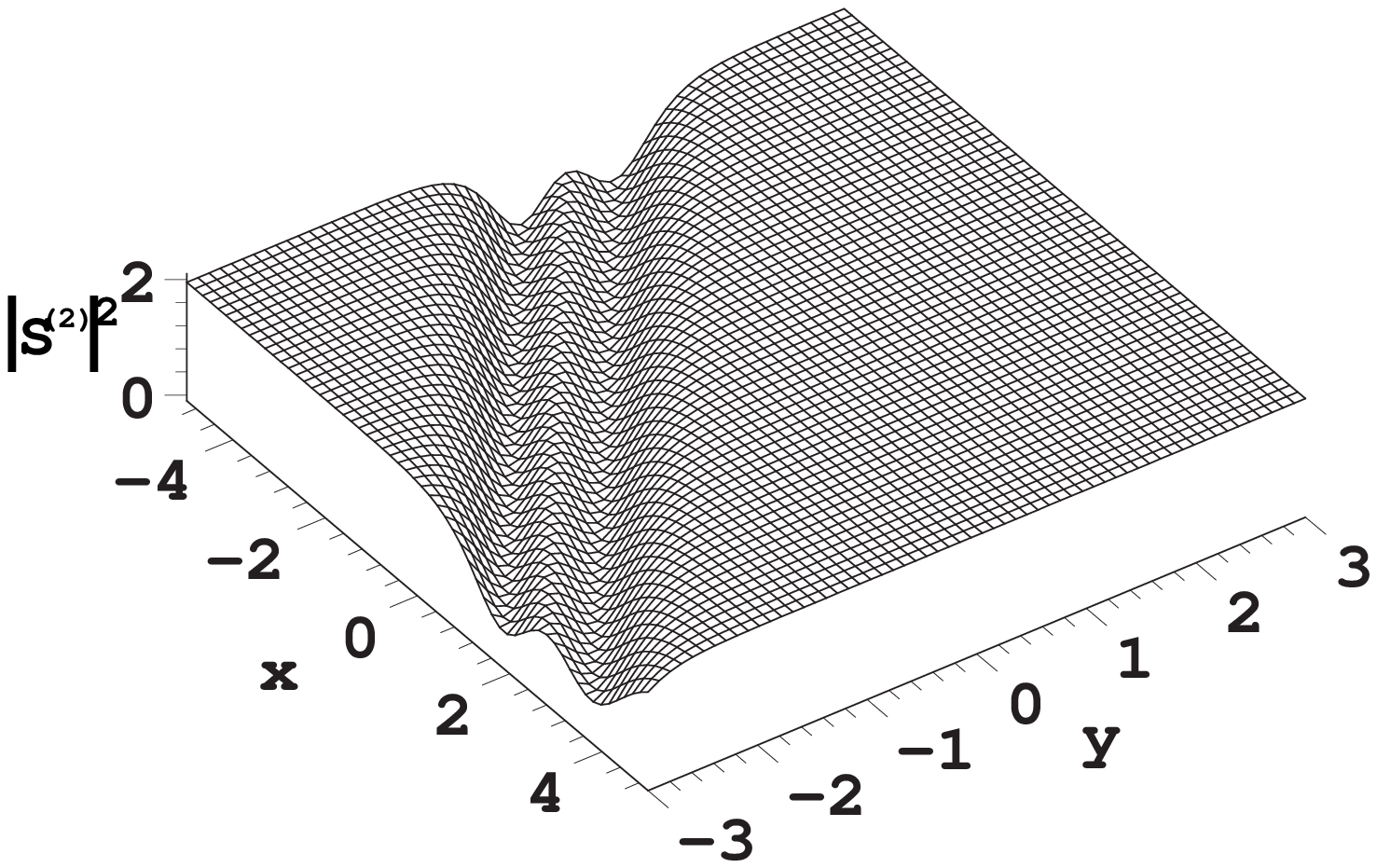}}\hspace{0.5cm}
\subfigure[]{\includegraphics[height=1.4in,width=1.9in]{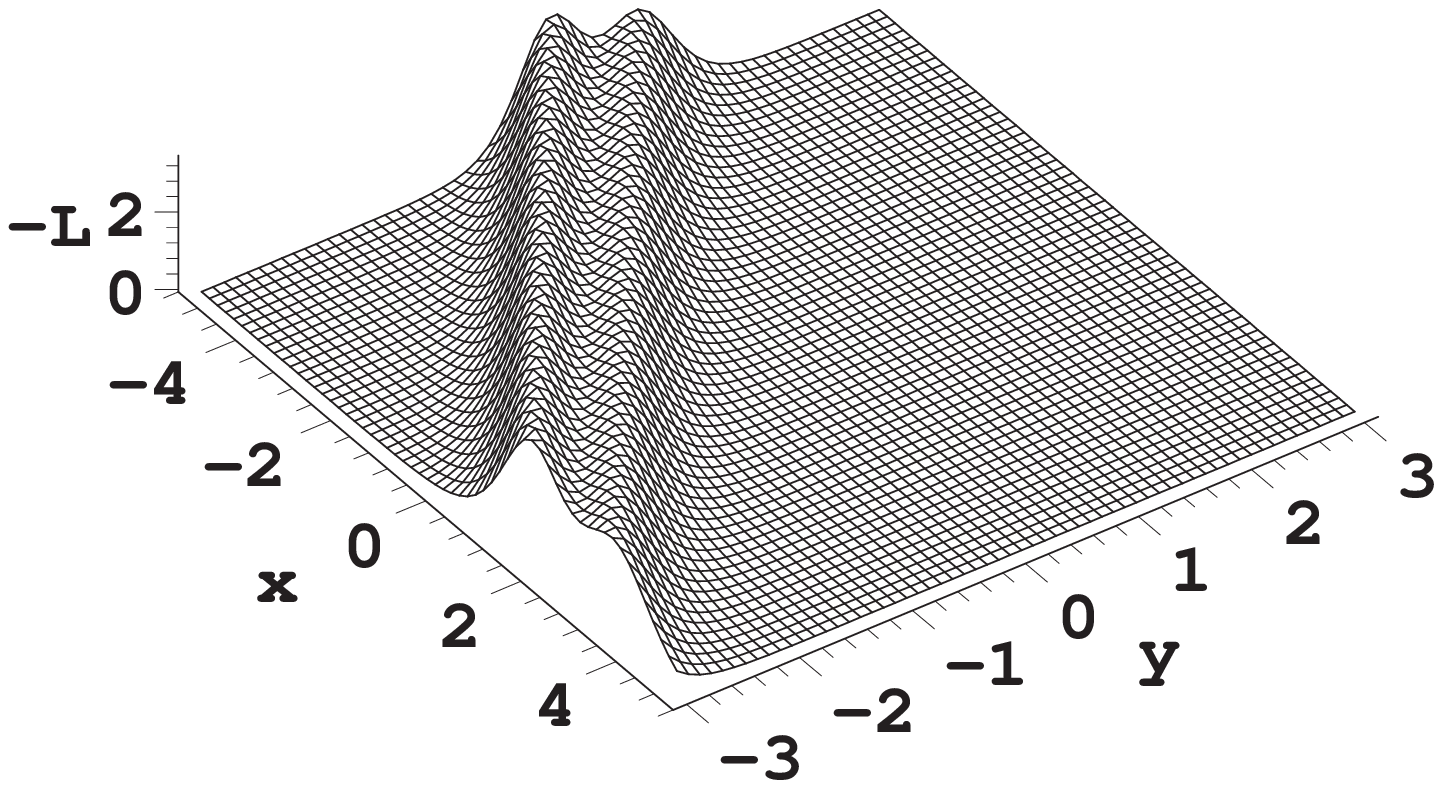}}
\caption{ The moving bound dark-dark soliton states with the parameters (\ref{nyo-92}) at time $t=-20$. }
\end{figure*}
\begin{figure*}[!htbp]
\centering
\subfigure[]{\includegraphics[height=1.4in,width=1.9in]{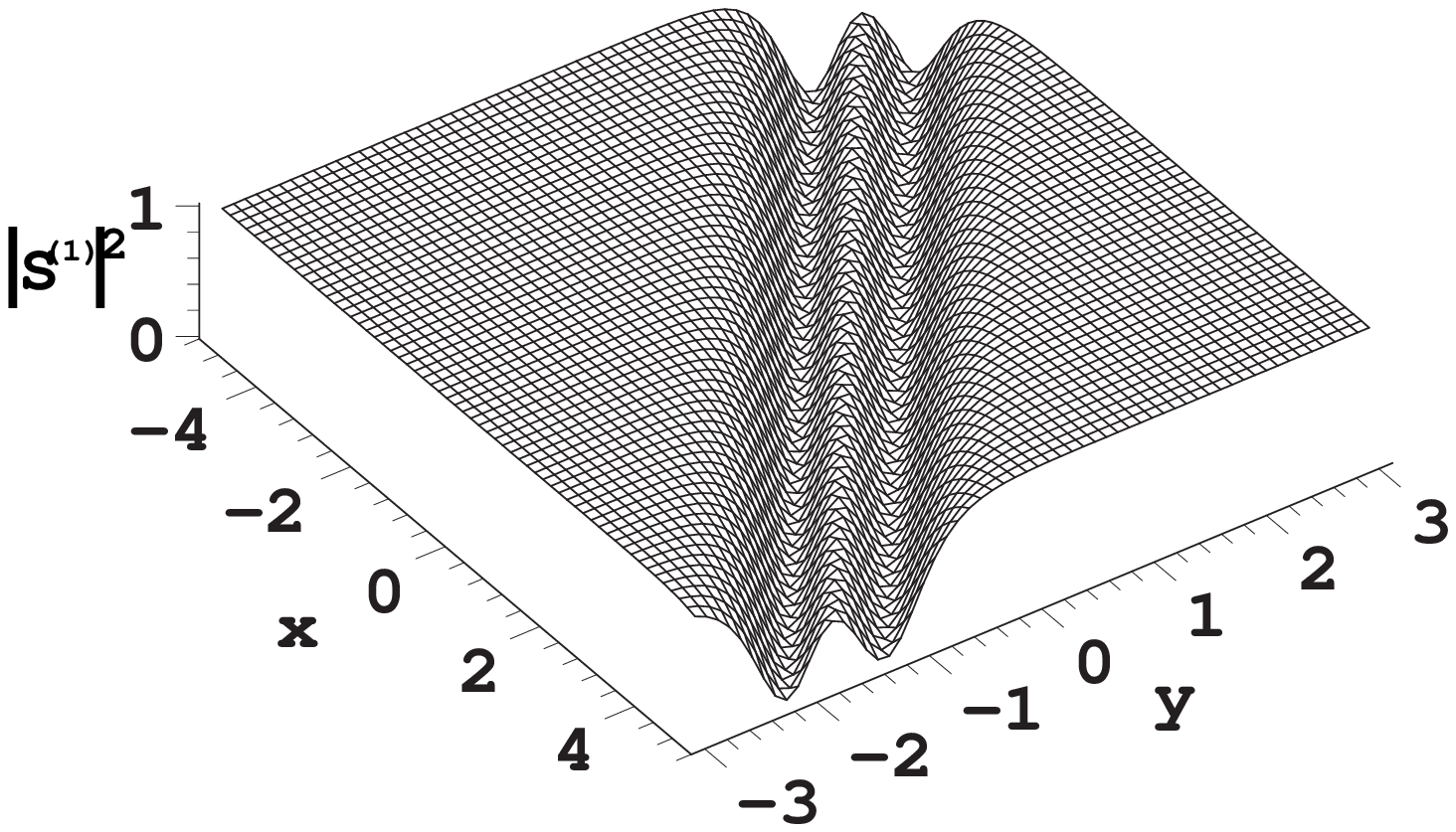}}\hspace{0.5cm}
\subfigure[]{\includegraphics[height=1.4in,width=1.9in]{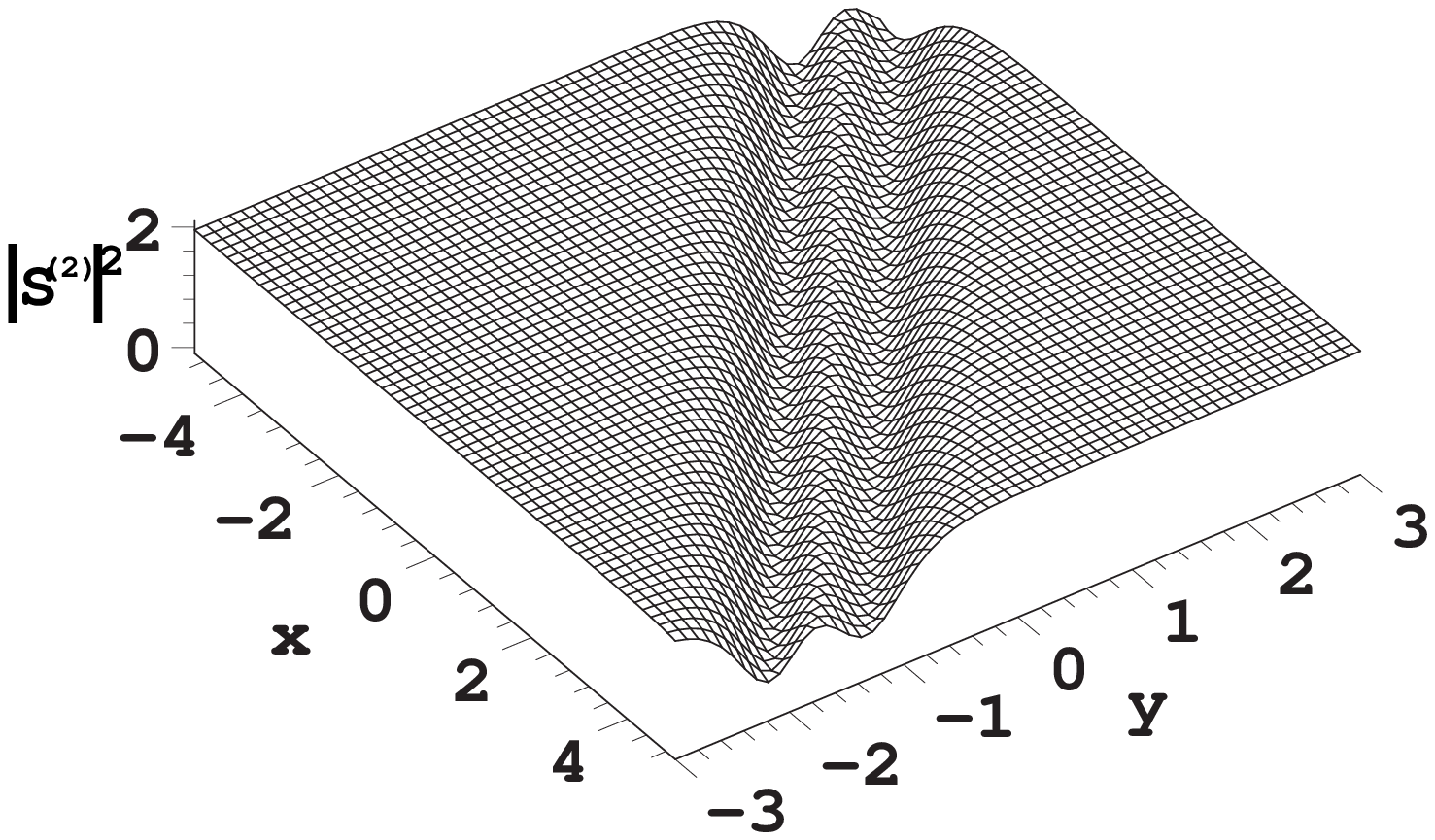}}\hspace{0.5cm}
\subfigure[]{\includegraphics[height=1.4in,width=1.9in]{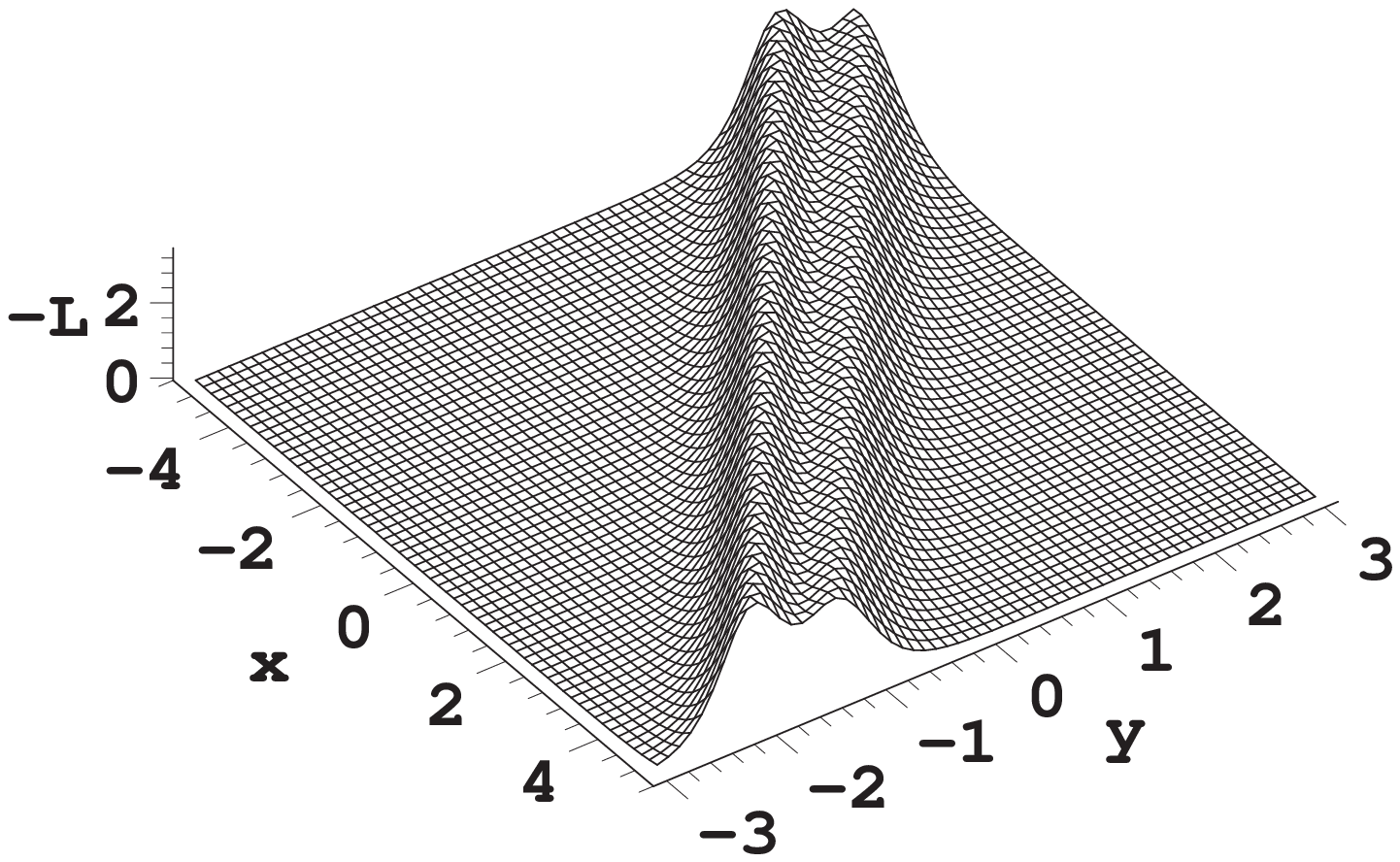}}
\caption{  The moving bound dark-dark soliton states with the parameters (\ref{nyo-92}) at time $t=0$. }
\end{figure*}
\begin{figure*}[!htbp]
\centering
\subfigure[]{\includegraphics[height=1.4in,width=1.9in]{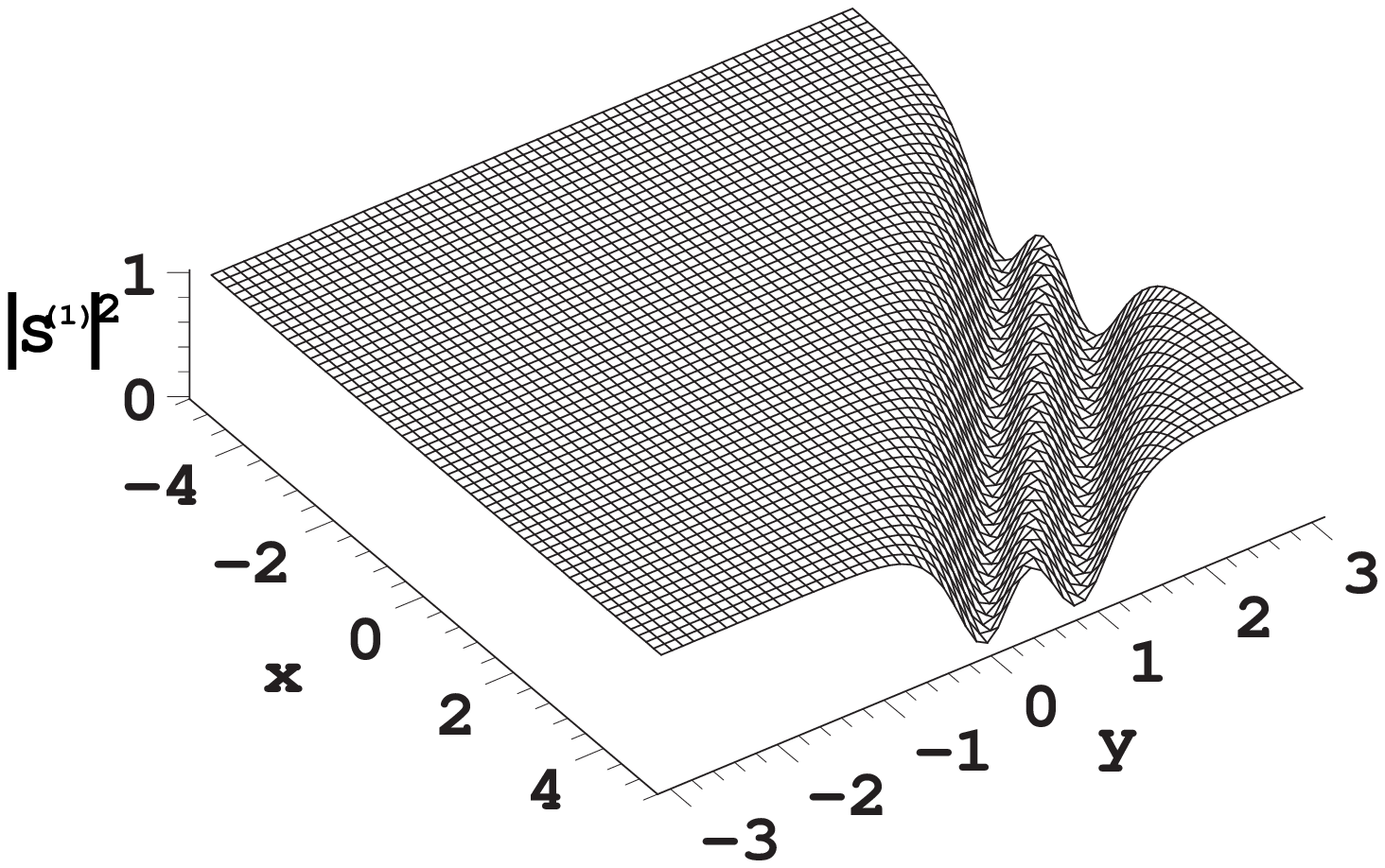}}\hspace{0.5cm}
\subfigure[]{\includegraphics[height=1.4in,width=1.9in]{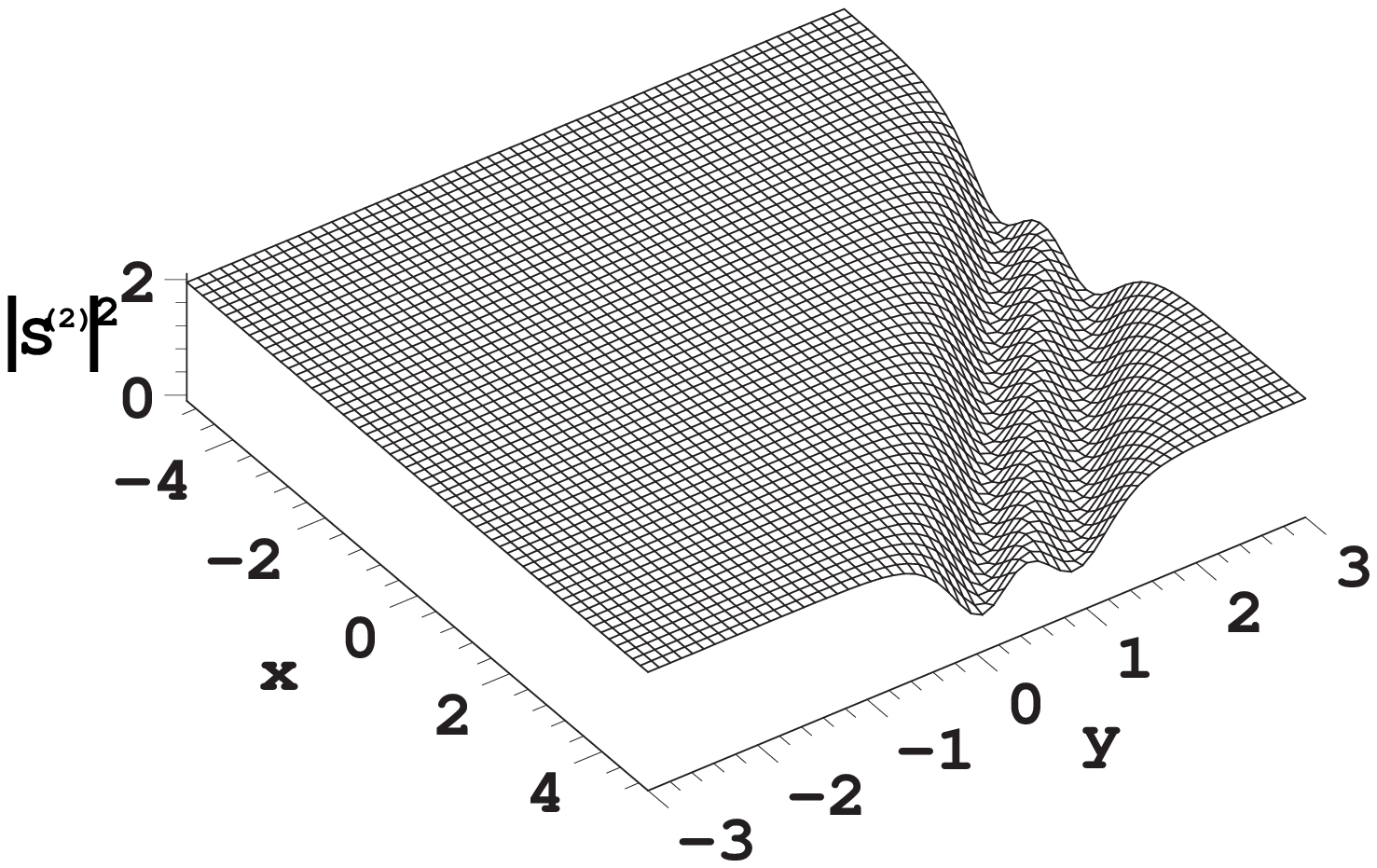}}\hspace{0.5cm}
\subfigure[]{\includegraphics[height=1.4in,width=1.9in]{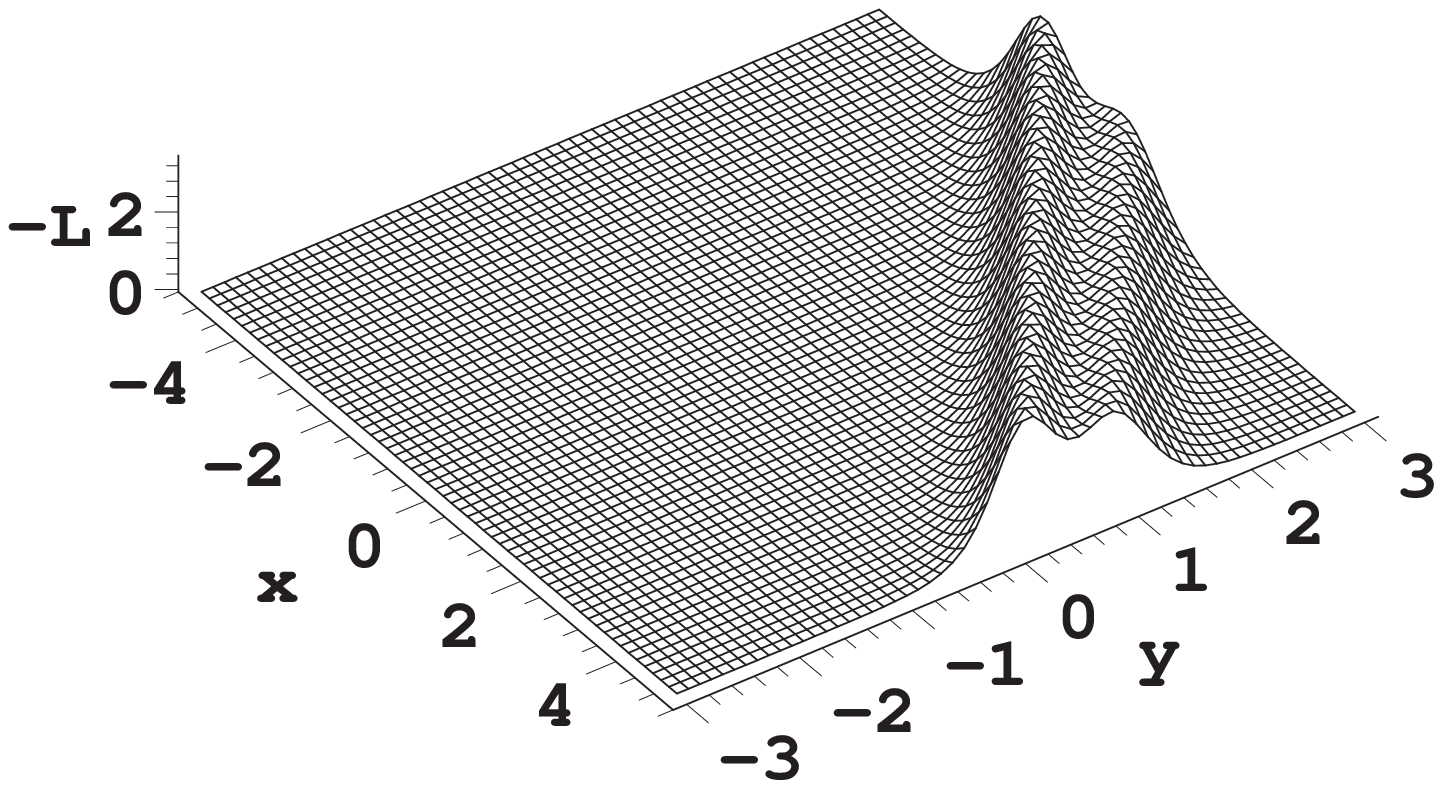}}
\caption{  The moving bound dark-dark soliton states with the parameters (\ref{nyo-92}) at time $t=20$. }
\end{figure*}

We should point out that, in both stationary and moving bound states,
the short wave components acquire non-zero phase shifts but the long
wave component has no phase shift as $x$ and $y$ vary from $-\infty$ to
$+\infty$. This feature is the same as the general two-dark-dark
solitons. Indeed, if $2\phi^{(1)}_j$ and $2\phi^{(2)}_j$ represent the
phases of complex constants $K^{(1)}_j$ and $K^{(2)}_j$ respectively,
the phase shifts for all components are $S^{(1)}_{\small  \textrm{phase
shift}}=2\phi^{(1)}_1+2\phi^{(1)}_2$, $S^{(2)}_{\small\textrm{phase
shift}}=2\phi^{(2)}_1+2\phi^{(2)}_2$ and $-L_{\small \textrm{phase shift}}=0$.
The total phase shifts of each short wave component are equal to the sum of the individual ones of the two dark solitons while the phase shifts of the long wave component are always zero.

Most recently, Sakkaravarthi and Kanna presented three bright-bright soltion bound states of the coupled
YO system\cite{sakkaravarthi2013dynamics}.
It is natural to see whether or not three- or higher-order dark-dark-soliton bound states exist in the coupled YO system. To ensure three- or higher-order dark-dark soliton bound states, at least three distinct values of $p_j$ should exist. For the stationary bound states, as two subcases stated in Sect. 5.1, $a_j$ can either take arbitrary positive value or is determined by $b_j$. So it is not difficult to construct the stationary dark-dark bound state up to arbitrary order. However, for the moving dark-dark soliton bound states, from (\ref{nyo-91}), all $b_j$'s values must be the same, which ends up at most two distinct values of $a_j$. This observation leads to a conclusion that there is no three- or higher-order moving bound states.


\section{General $N$ dark soliton solutions of the one- and two-dimensional multi-component YO systems}

As a matter of fact, we can extend our previous analysis to the 1D and
2D multi-component coupled YO systems. It is known that the multi-bright
soliton solutions can be derived from the reduction of the
multi-component KP hierarchy, whereas, the multi-dark soliton solutions
are obtained from the reduction of the single KP hierarchy but with
multiple copies of shifted singular points.
Therefore, the general dark soliton solutions for the multi-component YO
systems
can be constructed in the same spirit as the two-component YO system.
The details are omitted here, and we present only the results for both 1D and 2D multi-component YO systems.

To seek for $N$-dark soliton solutions, the 2D multi-component YO system
consisting of
$M$ short wave components and one long wave component
\begin{eqnarray}
\label{nyo-93} \hspace{-1.3cm}&&\textmd{i} (S^{(k)}_t+S^{(k)}_y)-S^{(k)}_{xx}+LS^{(k)}=0,\ \ k=1,2,\cdots,M\\
\label{nyo-94} \hspace{-1.3cm}&& L_t=2\sum^{M}_{k=1}\sigma_k|S^{(k)}|^2_x\,,
\end{eqnarray}
is transformed to the following bilinear form
\begin{eqnarray}
\label{nyo-97} \hspace{-1.3cm}&&
 [\textmd{i}(D_t+D_y-2\alpha_kD_x)-D^2_x]g_k \cdot f=0, \\
\hspace{-1.3cm}&&\qquad \qquad \qquad \qquad \qquad  k=1,2,\cdots,M\nonumber\\
\label{nyo-98} \hspace{-1.3cm}&& [D_tD_x-2\sum^{M}_{k=1}\sigma_k\rho^2_k]f \cdot f+2\sum^{M}_{k=1}\sigma_k\rho^2_kg_kg^*_k=0,
\end{eqnarray}
through the dependent variable transformations:
\begin{eqnarray}
\hspace{-1cm}&& S^{(k)}= \rho_k
 \textmd{e}^{\textmd{i}[\alpha_kx+\beta_k y
 -(\beta_k-\alpha^2_k)t+\zeta_{k0}]}\frac{g_k}{f},\ \ k=1,2,\cdots,M\nonumber\\
\hspace{-1cm}&&L= -2\left(\log f\right)_{xx}, \label{nyo-95}
\end{eqnarray}
where $\alpha_k,\beta_k, \rho_k,\zeta_{k0}$ are real constants.

Similar to the procedure discussed in Sect. 2.2, taking into account the Gram type determinant solutions of the KP hierarchy, one can obtain $N$-dark soliton solutions as follows:
\begin{eqnarray}
\label{nyo-99} \hspace{-1cm}&&f=\Bigg|  \delta_{ij} + \frac{1}{p_i+p^*_j} \textmd{e}^{\xi_i+\xi^*_j}  \Bigg|_{N\times N}, \\
\label{nyo-100}\hspace{-1cm} && g_k=\Bigg| \delta_{ij} + \left(-\frac{p_i-\textmd{i}\alpha_k}{p^*_j+\textmd{i}\alpha_k}\right) \frac{1}{p_i+p^*_j} \textmd{e}^{\xi_i+\xi^*_j} \Bigg|_{N\times N},\ \
\end{eqnarray}
with
\begin{eqnarray*}
\hspace{-0.8cm}&& \xi_j=p_jx -
 \left(\sum^M_{k=1}\frac{\sigma_k\rho^2_k}{p_j-\textmd{i}\alpha_k}
  +\textmd{i}p^2_j\right)y \\
\hspace{-0.8cm}&&\qquad \qquad + \sum^M_{k=1}\frac{\sigma_k\rho^2_k}{p_j-\textmd{i}\alpha_k}t  + \xi_{j0}\,,
\end{eqnarray*}
where $p_j$ and $\xi_{j0}$ are complex constants.

Starting from the Wronskian solution of the KP hierarchy, $N$-dark soliton solutions in the Wronskian form can be constructed in the same way in Sect. 2.3, which is of the following form
\begin{eqnarray}
\label{nyo-101} \hspace{-1.2cm}&&f=\frac{1}{\mathcal{G}}\left| \begin{array}{ccccc}
\varphi_1 & \partial_{x_1}\varphi_1 & \cdots  & \partial^{(N-1)}_{x_1}\varphi_1  \\
\varphi_2 & \partial_{x_1}\varphi_2 & \cdots  & \partial^{(N-1)}_{x_1}\varphi_2  \\
\vdots & \vdots & \cdots  & \vdots  \\
\varphi_N & \partial_{x_1}\varphi_N & \cdots  & \partial^{(N-1)}_{x_1}\varphi_N
\end{array} \right|, \\
\label{nyo-102} \hspace{-1.2cm}&&
g_k=\frac{1}{\mathcal{C}_k \mathcal{G}}\left| \begin{array}{ccccc}
\bar{\varphi}^{(k)}_1 & \partial_{x_1}\bar{\varphi}^{(k)}_1 & \cdots  & \partial^{(N-1)}_{x_1}\bar{\varphi}^{(k)}_1  \\
\bar{\varphi}^{(k)}_2 & \partial_{x_1}\bar{\varphi}^{(k)}_2 & \cdots  & \partial^{(N-1)}_{x_1}\bar{\varphi}^{(k)}_2  \\
\vdots & \vdots & \cdots  & \vdots  \\
\bar{\varphi}^{(k)}_N & \partial_{x_1}\bar{\varphi}^{(k)}_N & \cdots  & \partial^{(N-1)}_{x_1}\bar{\varphi}^{(k)}_N
\end{array} \right|, \ \
\end{eqnarray}
with
\begin{eqnarray*}
&& \mathcal{G}=\Delta(-p^*_1,-p^*_2,\cdots, -p^*_N) \\
&& \qquad \times \prod^N_{j=1} \bigg( \prod^{N}_{k=1, k\neq j} -\frac{p_j+p^*_k}{p^*_j-p^*_k}   \bigg) \exp(-\xi^*_j), \\
&& \mathcal{C}_k=\prod^N_{j=1}(-1)^N(p^*_j+\textmd{i}\alpha_k),
\end{eqnarray*}
and
\begin{eqnarray*}
\hspace{-0.8cm}&& \varphi_j=\exp(\xi_j)+\exp(-\xi^*_j),\\
\hspace{-0.8cm}&& \bar{\varphi}^{(k)}_j=(p_j-\textmd{i}\alpha_k)\exp(\xi_j)-(p^*_j+\textmd{i}\alpha_k)\exp(-\xi^*_j),\\
\hspace{-0.8cm}&& \xi_j=p_jx -
 \Big(\sum^M_{k=1}\frac{\sigma_k\rho^2_k}{p_j-\textmd{i}\alpha_k}
 +\textmd{i}p^2_j\Big)y \\
\hspace{-0.8cm}&&\qquad \qquad + \sum^M_{k=1}\frac{\sigma_k\rho^2_k}{p_j-\textmd{i}\alpha_k}t  + \xi_{j0}\,,
\end{eqnarray*}
where $p_j$ and $\xi_{j0}$ are complex constants.

By the similar procedure discussed in Sect. 3, $N$-dark soliton
solutions for 1D multi-component
YO system are provided with the same form as the one for two-dimensional
case. In other words, $N$-dark soliton solutions for 1D and 2D
integrable systems can be deduced simultaneously without reformulating
the problem.
To be more specific, the following bilinear form
\begin{eqnarray}
\label{nyo-105}\hspace{-1.3cm} && [\textmd{i}(D_t-2\alpha_kD_x)-D^2_x]g_k \cdot f=0, \ \ k=1,2,\cdots,M\\
\label{nyo-106}\hspace{-1.3cm} && [D_tD_x-2\sum^{M}_{k=1}\sigma_k\rho^2_k]f \cdot f+2\sum^{M}_{k=1}\sigma_k\rho^2_kg_kg^*_k=0,
\end{eqnarray}
is converted from the 1D multi-component YO system
\begin{eqnarray}
\label{nyo-103} \hspace{-0.8cm}&&\textmd{i} S^{(k)}_t-S^{(k)}_{xx}+LS^{(k)}=0,\ \
 k=1,2,\cdots, M\\
\label{nyo-104}\hspace{-0.8cm} && L_t=2\sum^{M}_{k=1}\sigma_k|S^{(k)}|^2_x.
\end{eqnarray}
through dependent variable transformations
\begin{eqnarray}
\label{nyo-107}\hspace{-0.8cm} && S^{(k)}= \rho_k
 \textmd{e}^{\textmd{i}[\alpha_kx+\alpha^2_k
 t+\zeta_{k0}]}\frac{g_k}{f}, \ \ k=1,2,\cdots,M\\
\label{nyo-108}\hspace{-0.8cm} && L= -2\frac{\partial^2}{\partial{x^2}}\log f,
\end{eqnarray}
where $\alpha_k, \rho_k,\zeta_{k0}$ are real constants.

It is shown that, by imposing the constraint conditions
\begin{eqnarray}
\label{nyo-109}
 \sum^M_{k=1}\frac{\sigma_k\rho^2_k}{|p_j-\textmd{i}\alpha_k|^2}   = - \textmd{i} (p_j-p^{*}_j)\,,\quad j=1,2 \cdots, N\,,
\end{eqnarray}
the terms associated with $D_y$ in (\ref{nyo-97})-(\ref{nyo-98}) are
dropped out, the bilinear equations for the 2D multi-component YO system
are reduced to (\ref{nyo-105})-(\ref{nyo-106}) for the 1D
case. Therefore, the $N$-dark soliton solution for the 1D
multi-component coupled YO system shares the same Gram determinant form
(\ref{nyo-99})-(\ref{nyo-100}) or Wronskian form (\ref{nyo-101})-(\ref{nyo-102}) except
the constraint conditions (\ref{nyo-109}).

\section{Conclusions}

We have constructed the general multi-dark soliton solutions
in both the 1D
and the 2D multi-component coupled YO systems and analyzed their
dynamical behaviors.
General multi-dark soliton solutions
for 2D multi-component soliton systems have never been reported in literature.

By using the classical Hirota bilinear method, we have presented the
$N$-dark-dark
soliton solutions with the implicit dispersion relation in the 2D coupled
YO system containing
two short wave component and one long wave component.
By virtue of the reduction method of the KP hierarchy, $N$-dark-dark soliton
solutions expressed by
Gram type and Wronski type determinants are derived and proved.
The process of obtaining $N$-dark-dark soliton solutions elucidates the
connections of the YO system with other integrable systems in the KP
hierarchy,
which will be helpful for the further study of these systems. By further
reduction,
we also provide the general $N$-dark-dark soliton solutions for the 1D
coupled YO system
in the same form as the one for the 2D coupled YO system except some
constraint conditions.
The similar form of general $N$-dark soliton solutions in the 1D and 2D
multi-component YO systems
are constructed by simply inserting more copies of the shifts of
singular points.

We have further investigated the dynamical behaviors of one and two
dark-dark solitons in the 2D coupled YO system with two short wave
components.
In contrast with bright-bright soliton solutions, it is shown that
dark-dark soliton collisions
are elastic and there is no energy exchange in two components of each
soliton.

Moreover, the dark-dark soliton bound states including the stationary
and moving ones are discussed.
For the stationary case, the bound states exist up to arbitrary order,
whereas, for the moving case, only two-soliton bound state is possible under the condition that
the coefficients of nonlinear terms have opposite signs.

\section*{Acknowledgments}

J.C. appreciates the support by the China Scholarship Council. The
project is supported by the Global Change Research Program of China
(No.2015CB953904), National Natural Science Foundation of China (Grant
No.11275072, 11435005 and 11428102), Research Fund for the Doctoral
Program of Higher Education of China (No. 20120076110024), The Network
Information Physics Calculation of basic research innovation research
group of China (Grant No. 61321064), Shanghai Collaborative Innovation
Center of Trustworthy Software for Internet of Things (Grant
No. ZF1213), Shanghai Minhang District talents of high level
scientific research project, and CREST, JST..

\appendix
\section{}

In this appendix, we present the proof of Lemma 2.1 in Sect. 2.2\cite{ohta2011general}.
Consider functions $\phi_i$ and $\psi_{i}$ which satisfy the following
differential and difference rules:
\begin{eqnarray}\label{nyo-a-01}
\nonumber && \partial_{x_2}\phi_i(k,l)=\partial^2_{x_1}\phi_i(k,l),\\
\nonumber && \partial_{x_{-1}}\phi_i(k,l)=\phi_i(k-1,l),\\
&& \phi_i(k+1,l)=(\partial_{x_1}-a)\phi_i(k,l),\\
\nonumber && \partial_{x_2}\psi_i(k,l)=-\partial^2_{x_1}\psi_i(k,l),\\
\nonumber && \partial_{x_{-1}}\psi_i(k,l)=-\psi_i(k+1,l),\\
\nonumber && \psi_i(k-1,l)=-(\partial_{x_1}+a)\psi_i(k,l).
\end{eqnarray}
Define
\[
m_{ij}(k,l)=c_{ij}+\int \phi_i(k,l)\psi_j(k,l)\ dx_1\,,
\]
and the $N\times N$ matirx $\mathbf{m}(k,l)=(m_{ij}(k,l))_{1\leq i,j\leq N}$.
Then one can easily verify that the matrix elements $m_{ij}(k,l)$ satisfy
\begin{eqnarray}\label{nyo-a-02}
\hspace{-0.8cm}\nonumber && \partial_{x_1} m_{ij}(k,l)=\phi_i(k,l)\psi_j(k,l),\\
\hspace{-0.8cm}\nonumber && \partial_{x_2} m_{ij}(k,l)\\
\hspace{-0.8cm}\nonumber &&\quad =\left(\partial_{x_1}\phi_i(k,l)\right)\psi_j(k,l)-\phi_i(k,l)\left(\partial_{x_1}\psi_j(k,l)\right),\\
\hspace{-0.8cm}&& \partial_{x_{-1}} m_{ij}(k,l)=-\phi_i(k-1,l)\psi_j(k+1,l),\\
\hspace{-0.8cm}\nonumber &&  m_{ij}(k+1,l)=m_{ij}(k,l)+\phi_i(k,l)\psi_j(k+1,l)\,.
\end{eqnarray}
The functions $\phi_i$, $\psi_{i}$ and the matrix elements
$m_{ij}(k,l)$ in Lemma 2.1 satisfy these relations.

Then with the help of (\ref{nyo-a-01}) and (\ref{nyo-a-02}), one can check that
the derivatives and shifts of the $\tau$ function are expressed by the
bordered
determinants as follows\cite{ohta2011general,hirota2004direct}:
\begin{eqnarray}
\nonumber &&\partial_{x_1}\tau(k,l)
=
\left| \begin{array}{ccccc}
\mathbf{m}(k,l) & \Phi(k,l)  \\
-\Psi(k,l) & 0
\end{array} \right|,
\end{eqnarray}
\begin{eqnarray}
\nonumber &&\partial^2_{x_1}\tau(k,l)
=
\left| \begin{array}{ccccc}
\mathbf{m}(k,l) & \partial_{x_1}\Phi(k,l)  \\
-\Psi(k,l) & 0
\end{array} \right|\\
\nonumber&& \qquad \qquad +
\left| \begin{array}{ccccc}
\mathbf{m}(k,l) & \Phi(k,l)  \\
-\partial_{x_1}\Psi(k,l) & 0
\end{array} \right|,
\end{eqnarray}
\begin{eqnarray}
\nonumber &&\partial_{x_2}\tau(k,l)
=
\left| \begin{array}{ccccc}
\mathbf{m}(k,l) & \partial_{x_1}\Phi(k,l)  \\
-\Psi(k,l) & 0
\end{array} \right|\\
\nonumber&& \qquad \qquad
-
\left| \begin{array}{ccccc}
\mathbf{m}(k,l) & \Phi(k,l)  \\
-\partial_{x_1}\Psi(k,l) & 0
\end{array} \right|,
\end{eqnarray}
\begin{eqnarray}
\nonumber &&\partial_{x_{-1}}\tau(k,l)
=
\left| \begin{array}{ccccc}
\mathbf{m}(k,l) & \Phi(k-1,l)  \\
\Psi(k+1,l) & 0
\end{array} \right|,
\end{eqnarray}
\begin{eqnarray}
\nonumber &&(\partial_{x_1}\partial_{x_{-1}}-1)\tau(k,l)\\
\nonumber&& \quad
=\left| \begin{array}{ccccc}
\mathbf{m}(k,l) & \Phi(k-1,l) & \Phi(k,l) \\
\Psi(k+1,l) & 0 & -1 \\
-\Psi(k,l) & -1 & 0
\end{array} \right|,
\end{eqnarray}
\begin{eqnarray}
\nonumber &&\tau(k+1,l)
=\left| \begin{array}{ccccc}
\mathbf{m}(k,l) & \Phi(k,l)  \\
-\Psi(k+1,l) & 1
\end{array} \right|,
\end{eqnarray}
\begin{eqnarray}
\nonumber &&\tau(k-1,l)
=
\left| \begin{array}{ccccc}
\mathbf{m}(k,l) & \Phi(k-1,l)  \\
\Psi(k,l) & 1
\end{array} \right|,
\end{eqnarray}
\begin{eqnarray}
\hspace{-0.8cm}\nonumber &&(\partial_{x_1}+a)\tau(k+1,l)
=
\left| \begin{array}{ccccc}
\mathbf{m}(k,l) & \partial_{x_1}\Phi(k,l)  \\
-\Psi(k+1,l) & a
\end{array} \right|,
\end{eqnarray}
\begin{eqnarray}
\hspace{-0.8cm}\nonumber &&(\partial_{x_1}+a)^2\tau(k+1,l)
=
\left| \begin{array}{ccccc}
\mathbf{m}(k,l) & \partial^2_{x_1}\Phi(k,l)  \\
-\Psi(k+1,l) & a^2
\end{array} \right|
\\
\hspace{-0.8cm}\nonumber &&\quad
+
\left| \begin{array}{ccccc}
\mathbf{m}(k,l) & \partial_{x_1}\Phi_i(k,l) & \Phi_i(k,l) \\
-\Psi(k+1,l) & a & 1 \\
-\Psi(k,l) & 0 & 0
\end{array} \right|,
\end{eqnarray}
\begin{eqnarray}
\hspace{-0.8cm}\nonumber &&(\partial_{x_2}+a^2)\tau(k+1,l)
=
\left| \begin{array}{ccccc}
\mathbf{m}(k,l) & \partial^2_{x_1}\Phi(k,l)  \\
-\Psi(k+1,l) & a^2
\end{array} \right|\\
\hspace{-0.8cm}\nonumber &&\quad -
\left| \begin{array}{ccccc}
\mathbf{m}(k,l) & \partial_{x_1}\Phi(k,l) & \Phi(k,l) \\
-\Psi(k+1,l) & a & 1 \\
-\Psi(k,l) & 0 & 0
\end{array} \right|,
\end{eqnarray}
where the bordered determinants are defined as
\begin{eqnarray}
\hspace{-0.8cm}\nonumber&&
\left| \begin{array}{ccccc}
\mathbf{m}(k,l) & \Phi(k,l)  \\
-\Psi(k,l) & 0
\end{array} \right|\\
\hspace{-0.8cm}\nonumber && \equiv
\small
\left| \begin{array}{ccccc}
m_{11}(k,l) & m_{12}(k,l) & \cdots & m_{1N}(k,l) & \phi_1(k,l) \\
m_{21}(k,l) & m_{22}(k,l) & \cdots & m_{2N}(k,l) & \phi_2(k,l) \\
\vdots & \vdots & \vdots & \vdots & \vdots \\
m_{N1}(k,l) & m_{N2}(k,l) & \cdots & m_{NN}(k,l) & \phi_N(k,l) \\
-\psi_1(k,l) & -\psi_2(k,l)  &\cdots & -\psi_N(k,l) & 0
\end{array} \right|,
\end{eqnarray}
and $\Phi(k,l)=(\phi_1(k,l),\phi_2(k,l),\cdots,
\phi_N(k,l))^{\mathrm{T}}$ and
$\Psi(k,l)=(\psi_1(k,l),\psi_2(k,l),\cdots,
\psi_N(k,l))$.

By using the above relations, one can verify
\begin{eqnarray}
\nonumber &&(D_{x_2}-D^2_{x_1}-2a D_{x_1})\tau(k+1,l)\cdot \tau(k,l)\\
\nonumber &&=-2\left|
\begin{array}{ccc}
\mathbf{m}(k,l) & \partial_{x_1}\Phi(k,l) & \Phi(k,l) \\
-\Psi(k+1,l) & a & 1 \\
-\Psi(k,l) & 0 & 0
\end{array} \right|\\
\nonumber &&\quad \times
\left| \begin{array}{c}
\mathbf{m}(k,l)
\end{array} \right|
\\
\nonumber &&+2
\left| \begin{array}{cc}
\mathbf{m}(k,l) & \partial_{x_1}\Phi(k,l)  \\
-\Psi(k+1,l) & 0
\end{array} \right|\\
\nonumber && \quad \times
\left| \begin{array}{cc}
\mathbf{m}(k,l) & \Phi(k,l)  \\
-\Psi(k,l) & 0
\end{array} \right|
\\
 &&-2
\left| \begin{array}{cc}
\mathbf{m}(k,l) & \Phi(k,l)  \\
-\Psi(k+1,l) & 0
\end{array} \right|\nonumber\\
&& \quad \times
\left| \begin{array}{cc}
\mathbf{m}(k,l) & \partial_{x_1}\Phi(k,l)  \\
-\Psi(k,l) & 0
\end{array} \right|,\label{nyo-a-03}
\end{eqnarray}
\begin{eqnarray}
\nonumber &&\left(\frac{1}{2}D_{x_1}D_{x_{-1}}-1\right)
\tau(k,l)\cdot \tau(k,l)+\tau(k+1,l)\tau(k-1,l)\\
\nonumber &&=\left| \begin{array}{ccccc}
\mathbf{m}(k,l) & \Phi(k-1,l) & \Phi(k,l) \\
\Psi(k+1,l) & 0 & -1 \\
-\Psi(k,l) & -1 & 0
\end{array} \right|\\
\nonumber &&\quad \times
\left| \begin{array}{ccccc}
\mathbf{m}(k,l)
\end{array} \right|
\\
\nonumber &&-
\left| \begin{array}{ccccc}
\mathbf{m}(k,l) & \Phi(k,l)  \\
-\Psi(k,l) & 0
\end{array} \right|\\
\nonumber && \quad \times
\left| \begin{array}{ccccc}
\mathbf{m}(k,l) & \Phi(k-1,l)  \\
\Psi(k+1,l) & 0
\end{array} \right|
\\
\nonumber&&+
\left| \begin{array}{ccccc}
\mathbf{m}(k,l) & \Phi(k,l)  \\
-\Psi(k+1,l) & 1
\end{array} \right|\\
&&\quad \times
\left| \begin{array}{ccccc}
\mathbf{m}(k,l) & \Phi(k-1,l)  \\
\Psi(k,l) & 1
\end{array} \right|. \label{nyo-a-04}
\end{eqnarray}
Both (\ref{nyo-a-03}) and (\ref{nyo-a-04}) are identically zero
because of the Jacobi identities\cite{hirota2004direct} and hence
$\tau(k,l),\tau(k+1,l)$ and $\tau(k-1,l)$ satisfy the bilinear equations (\ref{nyo-21}) and (\ref{nyo-22}).
In a similar way, one can prove the other two bilinear identities (\ref{nyo-23}) and (\ref{nyo-24}).

\section{}

Here we present the proof of Lemma 2.4 in Sect. 2.3.

Consider the $\tau$-function
\begin{eqnarray}
\nonumber \hspace{-0.8cm}&&\tau(k,l)\\
\hspace{-0.8cm}&&=\left| \begin{array}{ccccc}
\varphi_1(k,l) & \partial_{x_1}\varphi_1(k,l) & \cdots  & \partial^{(N-1)}_{x_1}\varphi_1(k,l)  \\
\varphi_2(k,l) & \partial_{x_1}\varphi_2(k,l) & \cdots  & \partial^{(N-1)}_{x_1}\varphi_2(k,l)  \\
\vdots & \vdots & \cdots  & \vdots  \\
\varphi_N(k,l) & \partial_{x_1}\varphi_N(k,l) & \cdots  & \partial^{(N-1)}_{x_1}\varphi_N(k,l)
\end{array}
\right|,\nonumber\\
\hspace{-0.8cm}&&
\end{eqnarray}
where functions $\varphi_i(k,l)$ satisfy the following linear dispersion relations:
\begin{eqnarray}
\hspace{-0.8cm}&& \partial_{x_1} \varphi^{(n)}_i(k,l)=\varphi^{(n)}_i(k+1,l)+a \varphi^{(n)}_i(k,l),\\
\hspace{-0.8cm}\nonumber  &&\partial_{x_2} \varphi^{(n)}_i(k,l)=\partial^2_{x_1} \varphi^{(n)}_i(k,l) \\
\hspace{-0.8cm} &&\quad =\varphi^{(n)}_i(k+2,l)+2a
 \varphi^{(n)}_i(k+1,l) +a^2 \varphi^{(n)}_i(k,l),\nonumber\\
\hspace{-0.8cm} &&\\
\hspace{-0.8cm} &&\partial_{x_{-1}} \varphi^{(n)}_i(k,l)=\varphi^{(n)}_i(k-1,l),\\
\hspace{-0.8cm} &&\partial_{x_1} \varphi^{(n)}_i(k,l)=\varphi^{(n)}_i(k,l+1)+b \varphi^{(n)}_i(k,l),\\
\hspace{-0.8cm}\nonumber  &&\partial_{x_2} \varphi^{(n)}_i(k,l)=\partial^2_{x_1} \varphi^{(n)}_i(k,l) \\
\hspace{-0.8cm} &&\quad =\varphi^{(n)}_i(k,l+2)+2b
 \varphi^{(n)}_i(k,l+1) +b^2 \varphi^{(n)}_i(k,l),\nonumber\\
\hspace{-0.8cm} &&\\
\hspace{-0.8cm} &&\partial_{y_{-1}} \varphi^{(n)}_i(k,l)=\varphi^{(n)}_i(k,l-1).
\end{eqnarray}
The functions $\varphi_i(k,l)$ in Lemma 2.4 satisfy these relations.

Let us introduce a simplified notation,
\begin{eqnarray}
\hspace{-0.8cm}\nonumber &&\left\vert
n_{k,l}, n+1_{k,l},  \ldots,  n+N-1_{k,l}
\right\vert\\
\hspace{-0.8cm}\nonumber &&
\equiv
\small
\left| \begin{array}{ccccc}
\partial_{x_1}^n\varphi_1(k,l) & \partial_{x_1}^{n+1}\varphi_1(k,l) & \cdots  & \partial_{x_1}^{n+N-1}\varphi_1(k,l)  \\
\partial_{x_1}^n\varphi_2(k,l) & \partial_{x_1}^{n+1}\varphi_2(k,l) & \cdots  & \partial_{x_1}^{n+N-1}\varphi_2(k,l)  \\
\vdots & \vdots & \cdots  & \vdots  \\
\partial_{x_1}^n\varphi_N(k,l) & \partial_{x_1}^{n+1}\varphi_N(k,l) & \cdots  & \partial_{x_1}^{n+N-1}\varphi_N(k,l)
\end{array} \right|.
\end{eqnarray}

One can rewrite the above $\tau$-function as
\begin{eqnarray}
\nonumber \hspace{-1cm}&&\tau(k,l)
=
\left\vert 0_{k,l}, 1_{k,l},  \ldots, N-2_{k,l}, N-1_{k,l}\right\vert
\\
\nonumber\hspace{-1cm} &&\quad =
\left\vert 0_{k,l}, 1_{k,l},  \ldots, N-2_{k,l}, (N-2_{k+1,l})+a(N-2_{k,l})\right\vert
\\
\nonumber\hspace{-1cm} &&\quad =
\left\vert 0_{k,l}, 1_{k,l},  \ldots, N-2_{k,l}, N-2_{k+1,l}\right\vert\\
\nonumber \hspace{-1cm}&&\qquad \cdots\\
\nonumber \hspace{-1cm}&&\quad =
\left\vert 0_{k,l}, 0_{k+1,l}, 1_{k+1,l},  \ldots , N-3_{k+1,l}, N-2_{k+1,l}\right\vert\\
\nonumber \hspace{-1cm}&& \qquad \cdots\\
\hspace{-1.3cm}&&\quad =
\left\vert 0_{k,l}, 0_{k+1,l}, 0_{k+2,l},  \ldots , 0_{k+N-2,l}, 0_{k+N-1,l}\right\vert.
\end{eqnarray}
For simplicity, we omit subscripts ${_{k,l}}$. Thus the above
$\tau$-function is written as
\begin{eqnarray*}
\tau(k,l)&=&\left\vert 0_{k,l}, 1_{k,l},  \ldots, N-2_{k,l},
	     N-1_{k,l}\right\vert\\
&=&\left\vert 0_{k,l}, 0_{k+1,l}, 0_{k+2,l},  \ldots , 0_{k+N-2,l}, 0_{k+N-1,l}\right\vert\\
&=&\left\vert 0, 1,  \ldots, N-2, N-1\right\vert.
\end{eqnarray*}

The differential formulas for $\tau(k,l)$ are derived as follows:
\begin{eqnarray}
\hspace{-1.2cm}\nonumber &&\partial_{x_1}\tau(k,l) = \partial_{x_1} \left\vert
0,  1  ,  \ldots ,  N-2,   N-1
\right\vert\\
\hspace{-1.2cm}&&\quad = \left\vert
0,  1  ,  \ldots ,  N-2,  N\right\vert
+Na\tau(k,l),\\
\hspace{-1.2cm}\nonumber && \partial^2_{x_1}\tau(k,l)
= \partial_{x_1} \left\vert
0,  1  ,  \ldots ,   N-2,   N
\right\vert+Na\partial_{x_1}\tau(k,l),\\
\hspace{-1.2cm}\nonumber &&\quad = \left\vert
0,  1  ,  \ldots ,  N-3,    N-1,   N
\right\vert\\
\hspace{-1.2cm}\nonumber  &&\qquad
+ \left\vert
0,  1  ,  \ldots ,  N-3,    N-2,   N+1
\right\vert\\
\hspace{-1.2cm}\nonumber  && \qquad +Na(\partial_{x_1}-Na)\tau(k,l) +Na\partial_{x_1}\tau(k,l)\\
\hspace{-1.2cm}\nonumber  &&\quad = \left\vert
0,  1,  \ldots ,  N-3,    N-1,   N
\right\vert\\
\hspace{-1.2cm}\nonumber  &&\qquad
+ \left\vert
0,  1,  \ldots , N-3,  N-2,   N+1
\right\vert\\
\hspace{-1.2cm}&& \qquad +2Na\partial_{x_1}\tau(k,l) -N^2a^2\tau(k,l),\\
\hspace{-1.2cm}\nonumber &&\partial_{x_2}\tau(k,l)= \partial_{x_2} \left\vert
0,  1  ,  \ldots ,   N-2,   N-1
\right\vert\\
\hspace{-1.2cm}\nonumber  &&\quad = -\left\vert
0,  1 ,  \ldots ,  N-3,    N-1,   N
\right\vert\\
\hspace{-1.2cm}\nonumber  &&\qquad
+ \left\vert
0,  1,  \ldots ,  N-3,  N-2,   N+1
\right\vert\\
\hspace{-1.2cm}&&\qquad  +2a\partial_{x_1}\tau(k,l) -Na^2\tau(k,l),\\
\nonumber \hspace{-1.2cm}&& \partial_{x_{-1}}\tau(k,l)= \partial_{x_{-1}} \left\vert
0,  1,  \ldots , N-2,   N-1
\right\vert\nonumber\\
\hspace{-1.2cm}&&
\quad =
\left\vert
-1,  1  ,  \ldots ,  N-2,   N-1
\right\vert,\\
\hspace{-1.2cm}\nonumber && \partial_{x_{-1}}\partial_{x_1}\tau(k,l)\\
\hspace{-1.2cm}\nonumber &&\quad
 = \partial_{x_{-1}} \left\vert
0,  1  ,  \ldots ,   N-2,   N
\right\vert+Na\partial_{x_{-1}}\tau(k,l)\\
\hspace{-1.2cm}\nonumber &&\quad =  \left\vert
-1,  1  ,  \ldots , N-2,   N
\right\vert
+ \left\vert
0,  1 ,  \ldots ,  N-2,   N-1
\right\vert\\
\hspace{-1.2cm}\nonumber  && \qquad + Na\partial_{x_{-1}}\tau(k,l)\\
\hspace{-1.2cm}  &&\quad =  \left\vert
-1,  1,  \ldots ,  N-2,  N
\right\vert+\tau(k,l)+ Na\partial_{x_{-1}}\tau(k,l).\nonumber\\
\end{eqnarray}

Consider the following determinant identities:
$$
\tiny
\left| \begin{array}{cccccccccccccc}
1 & \cdots & N-2 &\vbl4 & N-1 &\vbl4
& N & \vbl4 & N+1 &\vbl4
&  & & $\hbox{\O}$ &  \\
\multispan{14}\hblfil \\
 &  \hbox{\O}    &    &\vbl4 & N-1 &\vbl4 & N &\vbl4 & N+1 & \vbl4
& 0  & 1  &\cdots &N-2
\end{array} \right | = 0,
$$
and
$$\tiny
\left|
\begin{array}{cccccccccccccc}
-1 &\vbl4 & 1 & \cdots & N-2 &\vbl4 & N &\vbl4 & & &\hbox{\O} & &\vbl4 & N-1 \\
\multispan{14}\hblfil \\
-1 &\vbl4 & & \hbox{\O}    &    &\vbl4 & N &\vbl4 & 0
& 1 &\cdots & N-2 &\vbl4 & N-1
\end{array} \right | = 0.
$$
Applying the Laplace expansion to the left-hand side of these identities, we obtain the
Pl\"{u}cker relations
\begin{eqnarray}
\hspace{-0.8cm}\nonumber&&  \left\vert
1,  \ldots , N{-}2, N, N{+}1
\right\vert
\times
 \left\vert
0,  1 ,  \ldots ,  N{-}2,  N{-}1
\right\vert\\
\hspace{-0.8cm}\nonumber&&
- \left\vert
1 ,  \ldots , N{-}2,  N{-}1,  N{+}1
\right\vert
\times
 \left\vert
0,  1 ,  \ldots , N{-}2,   N
\right\vert\\
\hspace{-0.8cm}\nonumber&& +
 \left\vert
1 ,  \ldots ,  N{-}2, N{-}1,  N
\right\vert
\times
 \left\vert
0,  1 ,  \ldots ,  N{-}2,   N{+}1
\right\vert
=0,
\end{eqnarray}
\begin{eqnarray}
\hspace{-0.8cm}\nonumber&&
 \left\vert
{-}1,  1  ,  \ldots , N{-}2,  N
\right\vert
\times
 \left\vert
0,  1 ,  \ldots , N{-}2,  N{-}1
\right\vert\\
\hspace{-0.8cm}\nonumber&&
{-}
 \left\vert
{-}1,  1 ,  \ldots ,  N{-}2,  N{-}1
\right\vert
\times
 \left\vert
0,  1  ,  \ldots ,  N{-}2,   N
\right\vert
\\
\hspace{-0.8cm} \nonumber&& +
 \left\vert
1,  2,  \ldots ,N-2, N{-}1, N
\right\vert
\times
\left\vert
{-}1, 0, 1,  \ldots, N{-}3,  N{-}2
\right\vert=0\,.
\end{eqnarray}
By using the $\tau$-functions, these determinant identities are rewritten as
\begin{eqnarray*}
\small
&&
\frac{1}{2}\big(-\partial_{x_2}\tau(k+1,l)+\partial^2_{x_1}\tau(k+1,l)\\
&&\quad -2(N-1)a\partial_{x_1}\tau(k+1,l)
+N(N-1)a^2\tau(k+1,l)\big)\\
&&\quad \times\tau(k,l)\\
&&\quad -(\partial_{x_1}\tau(k+1,l)-Na\tau(k+1,l)) \\
&& \quad \times
 (\partial_{x_1}\tau(k,l)-Na \tau(k,l))\\
&&\quad +\tau(k+1,l)\\
&&\quad \times
\frac{1}{2}
\big(\partial_{x_2}\tau(k,l)+\partial^2_{x_1}\tau(k,l)-2(N+1)a\partial_{x_1}\tau(k,l)\\
&&\quad
+N(N+1)a^2\tau(k,l)\big)=0,\\
&& (\partial_{x_{-1}}\partial_{x_1}\tau(k,l)-\tau(k,l)-Na\partial_{x_{-1}}\tau(k,l))\times\tau(k,l)\\
&&\quad -\partial_{x_{-1}}\tau(k,l)\times
(\partial_{x_1}\tau(k,l)-Na\tau(k,l))\\
 && \quad
+\tau(k+1,l)\times\tau(k-1,l)=0,
\end{eqnarray*}
which are nothing but bilinear equations (\ref{nyo-21}) and (\ref{nyo-22}), respectively.
Eqs.(\ref{nyo-23}) and (\ref{nyo-24}) can be proved in a similar way.


\begin{thebibliography}{10}

\bibitem{ohta2007two}
Y. Ohta, K. Maruno and M. Oikawa,
J. Phys. A: Math. Theor. \textbf{40}, 7659 (2007).


\bibitem{oikawa1989two}
M. Oikawa, M. Okamura and M. Funakoshi,
 J. Phys. Soc. Japan \textbf{58}, 4416 (1989).

\bibitem{grimshaw1977modulation}
R.~H.~J. Grimshaw,
 Stud. Appl. Math. \textbf{56}, 241 (1977).


\bibitem{yajima1976formation}
N. Yajima and M. Oikawa,
 Prog. Theor. Phys. \textbf{56}, 1719 (1976).

\bibitem{benney1976}
D. J. Benney,
Stud. Appl. Math. \textbf{55}, 93 (1976).

\bibitem{djordjevic1977two}
V. D. Djordjevic and L. G. Redekopp,
 J. Fluid Mech. \textbf{79}, 703 (1977).

\bibitem{chowdhury2008long}
A. Chowdhury and J.~A. Tataronis,
 Phys. Rev. Lett. \textbf{100}, 153905 (2008).

\bibitem{ma1979some}
Y. C. Ma and L. G. Redekopp,
 Phys. Fluids \textbf{22}, 1872 (1979).

\bibitem{ma1978complete}
Y. C. Ma,
 Stud. Appl. Math. \textbf{59}, 201 (1978)

\bibitem{hirota2004direct}
R. Hirota,
\emph{The direct method in soliton theory}
(Cambridge University Press, Cambridge, 2004).

\bibitem{radha2005periodic}
R. Radha, C. S. Kumar, M. Lakshmanan, X. Y. Tang and S. Y. Lou,
 J. Phys. A: Math. Gen. \textbf{38}, 9649 (2005).

\bibitem{lai1999wave}
D. W. C. Lai and K. W. Chow,
 J. Phys. Soc. Japan \textbf{68}, 1847 (1999).

\bibitem{kanna2009higher}
T. Kanna, M. Vijayajayanthi, K. Sakkaravarthi and M. Lakshmanan,
 J. Phys. A: Math. Theor.  \textbf{42}, 115103 (2009).

\bibitem{kanna2014general}
T. Kanna, K. Sakkaravarthi and K. Tamilselvan,
 Phys. Rev. E  \textbf{88}, 062921 (2013).

\bibitem{radha2009collision}
R. Radha, C. S. Kumar, M. Lakshmanan and C. R. Gilson,
 J. Phys. A: Math. Theor. \textbf{42}, 102002 (2009).

\bibitem{kanna2012mixed}
T. Kanna, M. Vijayajayanthi and M. Lakshmanan,
 Phys. Rev. E \textbf{90}, 042901 (2014).

\bibitem{chow2013rogue}
K.~W. Chow, H.~N. Chan, D.~J. Kedziora and R.~H.~J. Grimshaw,
J. Phys. Soc. Japan \textbf{82}, 074001 (2013).

\bibitem{chen2014dark}
S. H. Chen, P. Grelu, and J. M. Soto-Crespo,
 Phys. Rev. E \textbf{89}, 011201 (2014).

\bibitem{ohta2011general}
Y. Ohta, D. S. Wang and J. Yang,
 Stud. Appl. Math. \textbf{127}, 345 (2011).

\bibitem{sakkaravarthi2013dynamics}
K. Sakkaravarthi and T. Kanna,
 Eur. Phys. J. Special Topics \textbf{222}, 641 (2013).

\bibitem{marunoohta2006}
K. Maruno and Y. Ohta,
 J. Phys. Soc. Japan \textbf{75}, 054002 (2006).

\bibitem{CK:08}
S. Chakravarty and Y. Kodama,
J. Phys. A: Math. Theor. \textbf{41} 275209 (2008).

\bibitem{CK:09}
S. Chakravarty and Y. Kodama,
Stud. Appl. Math. \textbf{123} 83 (2009).

\bibitem{K:10}
Y. Kodama,
J. Phys. A: Math. Theor. \textbf{43} 434004 (2010).

\bibitem{CLM:10}
S. Chakravarty, T. Lewkow and K. Maruno,
Appl. Anal. \textbf{89} 529 (2010).



\end{thebibliography}

\end{document}